\documentclass[twocolumn,reprint,amsmath,amssymb,aps]{revtex4-1}

\usepackage{lineno}
\usepackage{graphicx}% Include figure files
\usepackage{dcolumn}% Align table columns on decimal point
\usepackage{bm}% bold math
\usepackage{comment}
\usepackage{mathtools}
\usepackage{float}
\usepackage{bbm}
\usepackage{pgfplots}
\usepackage{xcolor}
\usepackage[all]{xy}
\usepgfplotslibrary{groupplots}
\pgfplotsset{width=10cm,compat=1.8}
\usepgfplotslibrary{fillbetween}
\begin{document}
\setlength{\abovedisplayskip}{10pt}
\setlength{\belowdisplayskip}{10pt}
\title{Understanding the transition from paroxysmal to persistent atrial fibrillation \\ from micro-anatomical re-entry in a simple model}% Force line breaks with \\

\author{Alberto Ciacci$^{1,2,4,*}$}
\author{Max Falkenberg$^{1,2,4,*}$}%
\author{Kishan A. Manani$^{1,2,3}$}%
\author{Tim S. Evans$^{1,2}$}%
\author{Nicholas S. Peters$^{4}$}
\author{Kim Christensen$^{1,2,4}$}%
 \affiliation{%
 	$^{1}$Blackett Laboratory, Imperial College London, London SW7 2BW, United Kingdom
 }%
 \affiliation{%
	$^{2}$Center for Complexity Science, Imperial College London, London SW7 2AZ, United Kingdom
 }%
 \affiliation{
	$^{3}$National Heart and Lung Institute, Imperial College London, London W12 0NN, United Kingdom
 }
 \affiliation{
	$^{4}$ElectroCardioMaths Programme, Imperial Centre for Cardiac Engineering, Imperial College London, London W12 0NN, United Kingdom
 }
 \affiliation{
	$^{*}$\textbf{These authors have equally contributed to this work}
}
\affiliation{
	\textbf{Physics corresponding author: kim.christensen@imperial.ac.uk}
	\textbf{Medical corresponding author: n.peters@imperial.ac.uk}
}

\date{\today}% It is always \today, today,

\begin{abstract}
    Atrial fibrillation (AF) is the most common cardiac arrhytmia, characterised by the chaotic motion of electrical wavefronts in the atria. In clinical practice, AF is classified under two primary categories: paroxysmal AF, short intermittent episodes separated by periods of normal electrical activity, and persistent AF, longer uninterrupted episodes of chaotic electrical activity. However, the precise reasons why AF in a given patient is paroxysmal or persistent is poorly understood. Recently, we have introduced the percolation based Christensen-Manani-Peters (CMP) model of AF which naturally exhibits both paroxysmal and persistent AF, but precisely how these differences emerge in the model is unclear. In this paper, we dissect the CMP model to identify the cause of these different AF classifications. Starting from a mean-field model where we describe AF as a simple birth-death process, we add layers of complexity to the model and show that persistent AF arises from re-entrant circuits which exhibit an asymmetry in their probability of activation relative to deactivation. As a result, different simulations generated at identical model parameters can exhibit fibrillatory episodes spanning several orders of magnitude from a few seconds to months. These findings demonstrate that diverse, complex fibrillatory dynamics can emerge from very simple dynamics in models of AF.
\end{abstract}

\pacs{Valid PACS appear here}% PACS, the Physics and Astronomy
\maketitle

%\tableofcontents

\section{\label{Introduction} Introduction}
\noindent

Atrial fibrillation (AF) is the most common cardiac arrhythmia with a growing prevalence worldwide \cite{patel2018}. It is characterised by the rapid, irregular beating of the atria, caused by the chaotic motion of electrical wavefronts. This lack of coordinated contraction may allow blood to clot, making AF the leading cause of ischaemic stroke in the over 75s \cite{hart2001}.

Despite over one hundred years of extensive research, the mechanisms underlying the initiation and maintenance of AF are still poorly understood \cite{nattel2002,schotten2016,federov2018,nattel2017B,Mann2018,calkins2017}. There are numerous controversies and conflicts in AF research, primary of which is the question of whether AF is driven and sustained by local (spatially fixed) sources of new fibrillatory waves, or whether AF is self-sustaining from the interaction and fragmentation of multiple meandering electrical wavelets in the atria \cite{schotten2016,nattel2017B,calkins2017}. Although this dispute is yet to be resolved, recent evidence appears to strengthen the case for local drivers as the primary mechanism of AF \cite{shivkumar2012,lee2015,narayan2013,haissaguerre2014,narayan2017,hansen2015,hansen2016,csepe2017,federov2018,baykaner2018,falkenberg2018}. 

Questions concerning the underlying mechanism of AF are of particular importance because they inform potential treatment strategies. Historically, treatment for AF has focused on mitigating potential symptoms and lowering the risk of stroke through the use of rate control, and anti-arrhythmic drugs \cite{atrial2002}. However, these treatments do not ``cure'' AF. Surgical ablation strategies have been developed to destroy, or isolate, the regions of atrial muscle thought to be responsible for initiating and sustaining AF \cite{calkins2017}. If local drivers are responsible for AF, ablating the focus of these drivers may terminate and prevent AF. If meandering wavelets underlie AF, ablation strategies which minimise the space wavelets can move into may be preferable. Although the leading ablation strategy, pulmonary vein isolation \cite{haissaguerre1998}, has a success rate of around 60\%, ablation still fails in a large subset of patients and AF re-occurs in many patients who were initially free of AF after surgery.

One of the key factors determining the likelihood of ablation success is the fraction of time a patient spends in AF \cite{calkins2017}. Clinically, AF is defined as paroxysmal if episodes are short and self-terminating. Conversely, long, uninterrupted AF episodes are referred to as persistent. In general, patients are much more likely to be free of AF after ablative treatment if AF is paroxysmal. The success rate is around 60\% for paroxysmal AF while it is 40\% for persistent AF after a three year follow-up \cite{tutuianu2015}. Recurrence rates are also significantly higher for persistent AF after an initially successful treatment. However, why a patient exhibits paroxysmal or persistent AF is unclear. In many cases paroxysmal AF will develop into persistent AF, but reversion to paroxysmal AF after years of persistent AF has also been observed \cite{sugihara2015}. Additionally, of the patients who initially exhibit paroxysmal AF, many develop persistent AF rapidly (after a few months), but others do not progress at all over several years \cite{kottkamp2013}.

The progression of AF from paroxysmal to persistent is often associated with the idea that ``AF begets AF'', most notably in the goat model \cite{wijffels1995}, but also with some evidence in human AF \cite{walters2016}. During AF, the atria undergo electrophysiological and structural changes which promote the progression of AF. Amongst these changes, the accumulation of fibrosis is a key factor in determining a patient's susceptibility to AF \cite{burstein2008,Friedrichs2012, platonov2011}. Fibrosis is also critical for the formation of re-entrant circuits that drive AF \cite{hansen2015,csepe2017,Friedrichs2012}. The emergence of a re-entrant circuit begins when the regular propagation of electrical wavefronts is disrupted by unidirectional blocks. These blocks leave an opening for the conduction to re-enter back from adjacent muscle fibres \cite{hansen2015,christensen2015}. When the atria accumulate fibrosis, the distribution of gap junctions between fibres becomes highly anisotropic, that is, adjacent fibres become less and less coupled. In this scenario, the re-entering conduction is less likely to be obstructed by refractory atrial muscle cells (myocites), finding the appropriate conditions for initiating a spatially-stable circuital conduction (i.e., a re-entrant circuit) which drives AF. However, the relationship between the absolute fibrosis burden in the atria and the persistence of AF is not clear -- two patients with an equivalent fibrosis burden may have drastically different heart rhythms (e.g. sinus rhythm vs. paroxysmal AF vs. persistent AF) \cite{kottkamp2013}. 
%Understanding how the atrial microstructure determines the persistence of AF may be key if we are to improve long term ablation success rates, particularly for patients with persistent AF.

In this paper, our aim is to better understand the relationship between AF persistence and the atrial microstructure using computational modelling. Computer models are a well established tool in cardiac electrophysiology, allowing for a range of experimental investigations that are not possible in a clinical, or laboratory setting. There are a wide variety of model types pitched at different scales and levels of complexity \cite{clayton2011}. Highly detailed, biophysical models focus on precisely modelling the exchange of ions across cardiomyocyte gap junctions to study the propagation of action potentials across topologically realistic cardiac tissue. However, the resolution of these models is often not ideal and they typically assume continuous cardiac tissue. Conversely, simplified discrete models focus on understanding the microstructure of cardiac tissue and how this effects the propagation of electrical wavefronts. The former are typically preferable when studying what effect a prospective drug might have on AF \cite{clayton2011,Niederer2019}, whereas the latter are most often used to study the effect of discontinuous tissue that might arise from the accumulation of fibrosis \cite{christensen2015,gokhale2017,makowiec2019}. The latter also have the benefit that their simplicity allows for much larger simulations suited to statistical analysis \cite{falkenberg2018,McGillivray2018,lin2017}, both in the duration of individual simulations and the resolution of phase spaces which can be generated. 

Previously, we have introduced the Christensen-Manani-Peters (CMP) model of AF, a simple percolation based model that investigates how the formation of re-entrant circuits is dependent on the decoupling of neighbouring muscle fibres, through the action of fibrosis or otherwise \cite{christensen2015}. The model is not a fully realistic representation of the atria and it does not consider the precise evolution and propagation of action potentials across the atrial tissue. However, the model effectively demonstrates from basic principles how re-entrant circuits can form if fibrosis accumulates in sufficient quantities in a given local area. Additionally, adaptations of the CMP model to 3D \cite{falkenberg2018} and to a realistic atrial topology based on a sheep heart \cite{falkenberg2019,zhao2012} have been successful at explaining a number of key clinical results and have generated a number of new hypotheses. This includes the distribution of re-entrant circuits in the atria, notably in the pulmonary vein sleeves and the atrial appendages, the appearance of re-entrant circuits as both re-entrant and focal sources, and the increased probability of ablation failure as AF becomes more persistent. Machine learning has been applied to the model to test prospective methods for automated re-entrant circuit detection from electrogram data \cite{McGillivray2018}, and other models inspired by the CMP approach have been used to study the heart rhythm of patients following a heart transplant \cite{makowiec2019}.

Consistent with clinical knowledge, the CMP model has shown that two tissues with the same total fibrosis burden may exhibit very different forms of AF \cite{manani2016} -- at the same level of coupling, different simulations may exhibit sinus rhythm, paroxysmal AF, persistent AF, or persistent AF before reverting to paroxysmal AF, see section~\ref{FUNCTIONAL}. This is because the formation of re-entrant circuits appears to be dependent on the local distribution of fibrosis, not the total fibrosis burden across the atria \cite{manani2016} -- this is in line with other computational studies on the effect of fibrosis on AF persistence \cite{zahid2016}. Despite these intriguing results, it is so far unclear how the variability in AF persistence arises from the specific processes taking place at the microscopic scale in the CMP model. Hence, the aim of this paper is to dissect the CMP model into its constituent parts to understand which parts of the model microstructure are responsible for the progression from paroxysmal to persistent AF.

A detailed overview of the CMP model will be given in Section \ref{CMP}, however, the key constituent elements include the lattice representing the atrial tissue, nodes representing individual muscle cells (or a block of cells), locations susceptible to unidirectional conduction block (where the propagating signal has a small probability of extinguishing), and lattice bonds representing the electrical connections between neighbouring nodes. The re-entrant circuits that form in the CMP model are spatially stable, but temporally intermittent -- they can turn on and off as a result of local conduction blocks. This has similarities to the self-regenerating renewal process proposed by others to explain cardiac fibrillation, where fibrillation is driven by the continuous birth and death of temporally intermittent drivers \cite{dharmaprani2019}. 

To dissect the CMP model, we first remove all spatial elements of the model. We do this by deriving a mean-field (MF) model where AF is described by a set of particles, representing critical structures which, when active, correspond to re-entrant circuits, evolving as a simple birth-death process. Our results indicate that the MF model significantly underestimates the probability of inducing AF relative to the CMP model, and that the MF model does not explain the emergence of persistent AF. 

At a second level of abstraction, we reintroduce the spatial components of the model, but carefully control the re-entrant circuits that form by inhibiting the interaction of multiple successive conduction blocks (within the same activation cycle). Like the MF model, this controlled version of the CMP model (cCMP) also underestimates the probability of inducing AF and the time in AF. However, the spatial elements of the cCMP model do not appear to make a difference to the absolute time spent in AF relative to the MF model. Only very small differences in the time in AF are observed between the MF and cCMP models, explained by small differences in the duration of individual fibrillatory events.

Finally, we show that the difference in the probability of inducing AF and the persistence of AF between the cCMP and CMP models can be explained by a series of complex re-entrant circuits that exhibit an assymetry between the probability of activating and deactivating. These circuits have a special property that they require fewer successive conduction blocks to initiate fibrillation than are needed to terminate fibrillation. We also demonstrate that in some cases several of these structures are coupled together such that the termination of one re-entrant circuit immediately activates a dormant neighbouring structure. These mechanisms result in a spectrum of individual fibrillatory events spanning several orders of magnitude from seconds to months. 

In the remainder of the paper, we outline the CMP model and review key results including previous work on the persistence of AF. Subsequently, we introduce the MF model and the cCMP model and explain why both these models underestimate the time spent in AF and the persistence of AF relative to the original CMP model. Finally, we put the CMP model and our results into a wider context and discuss their potential clinical impact, the limitations of our approach, and outline proposals for future work.

\section{\label{CMP} The CMP Model}
\subsection{Model Definition}
\noindent
The atrial muscle consists of tubiform cells (myocytes) of length $\Delta x^{'} \approx 100\mu\text{m}$ and diameter $\Delta y^{'} = \Delta z^{'} \approx 20 \mu\text{m}$ \cite{luke1991,verheule2003}. Myocytes are mainly connected longitudinally, composing discrete fibres that sporadically connect transversally. The Christensen-Manani-Peters (CMP) model condenses this branching network of anisotropic cells into an $L \times L$ square lattice of nodes \cite{christensen2015}. A node represents a single (or multiple) atrial cell(s). Nodes are longitudinally connected to their neighbours with probability $\nu_{\parallel} = 1$ and transversally with probability $0 \leq \nu_{\perp} \leq 1$. This creates long arrangements of nodes, mimicking the protracted, interlaced fibres in the atrium. This simplified representation of the myocardial architecture captures the anisotropic distribution of gap junctions \cite{luke1991}. Furthermore, it reproduces the dynamics of electrical impulses which mainly propagate longitudinally (along single muscle fibres) rather than transversally (across multiple fibres) \cite{McGillivray2018}. A cylindrical topology is obtained by applying open boundary conditions longitudinally and periodic boundary conditions transversally.

\begin{figure}
%\centering
%\includegraphics[width=\linewidth]{plot0.pdf}
\includegraphics[width=8.0cm,height = 7.0cm]{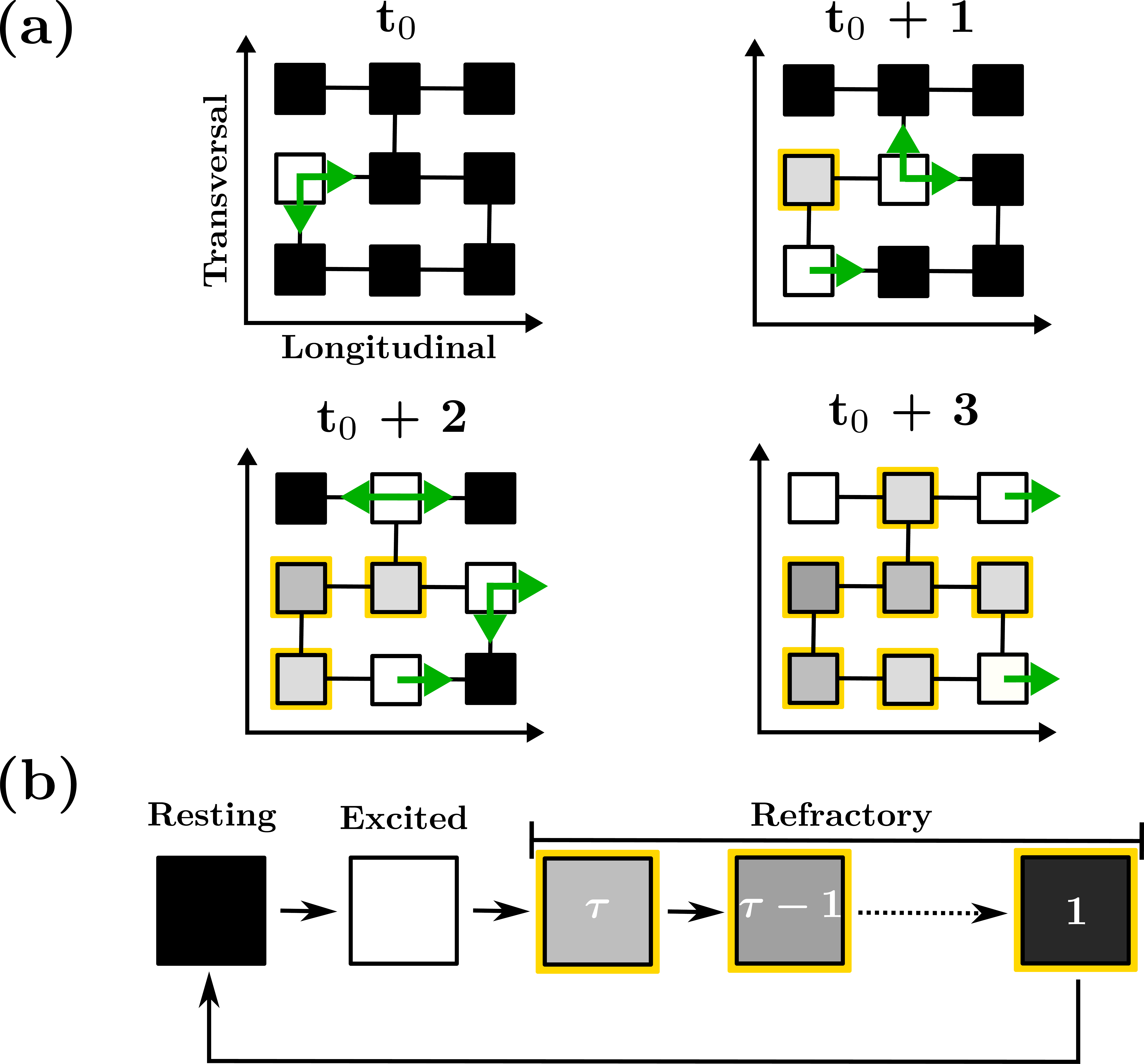}
%\end{figure}
\caption{(a) Propagation of the wave of excitation across a small region of the CMP lattice. Nodes are connected longitudinally with probability $\nu_{\parallel} = 1$ and transversally with probability $0 \leq \nu_{\perp} \leq 1$ . Excited nodes (white squares) continue the propagation of the wavefront by activating their neighbouring resting nodes (black squares) before entering into a refractory state (grey scale squares with yellow borders) for the next $\tau$ time steps. Depending on the architecture of the region, the excitation can proceed forward, backward and across fibres. (b) The full progression of a node through the three states of the electrical cycle: resting (black), excited (white) and refractory (grey scale with yellow borders).\label{first}}
\end{figure}

Nodes follow a well defined electrical cycle characterized by three different states: resting (a node that can be excited), excited or refractory (after exciting, the node cannot be excited for the next $\tau$ time steps). This course mimics the membrane potential of real myocardial cells.  At a given time $t$, an excited node prompts the neighbouring resting nodes to become excited at time $t + 1$. An excited node at time $t$ enters into a refractory state at time $t + 1$. The duration of the refractory period is $\tau$ time steps, see Fig.~\ref{first}.

In the CMP model, a fraction $\delta$ of nodes are susceptible to conduction block. These nodes are identified at the beginning of a simulation and are fixed in space. The probability that nodes that are susceptible to conduction block fail to excite is arbitrarily set to $\epsilon = 0.05$; the effect of varying this parameter is discussed in section \ref{MB}. This probability of failure refers to the probability that a node susceptible to conduction block will not excite when prompted to do so by a neighbouring active node. This leaves us with a very simple framework in which the fraction of transversal connections, $\nu_{\perp}$, and the fraction of nodes that are susceptible to conduction block, $\delta$, serve as control parameters. For simplicity, we set $\delta = 0.01$ and examine how the behavior of the system varies with $\nu_{\perp}$. The effect of changing $\delta$ is demonstrated in section~\ref{FUNCTIONAL} and has been investigated in \cite{kishanthesis}.

\begin{figure}
\centering
\includegraphics[width=5.0cm, height = 10.0cm]{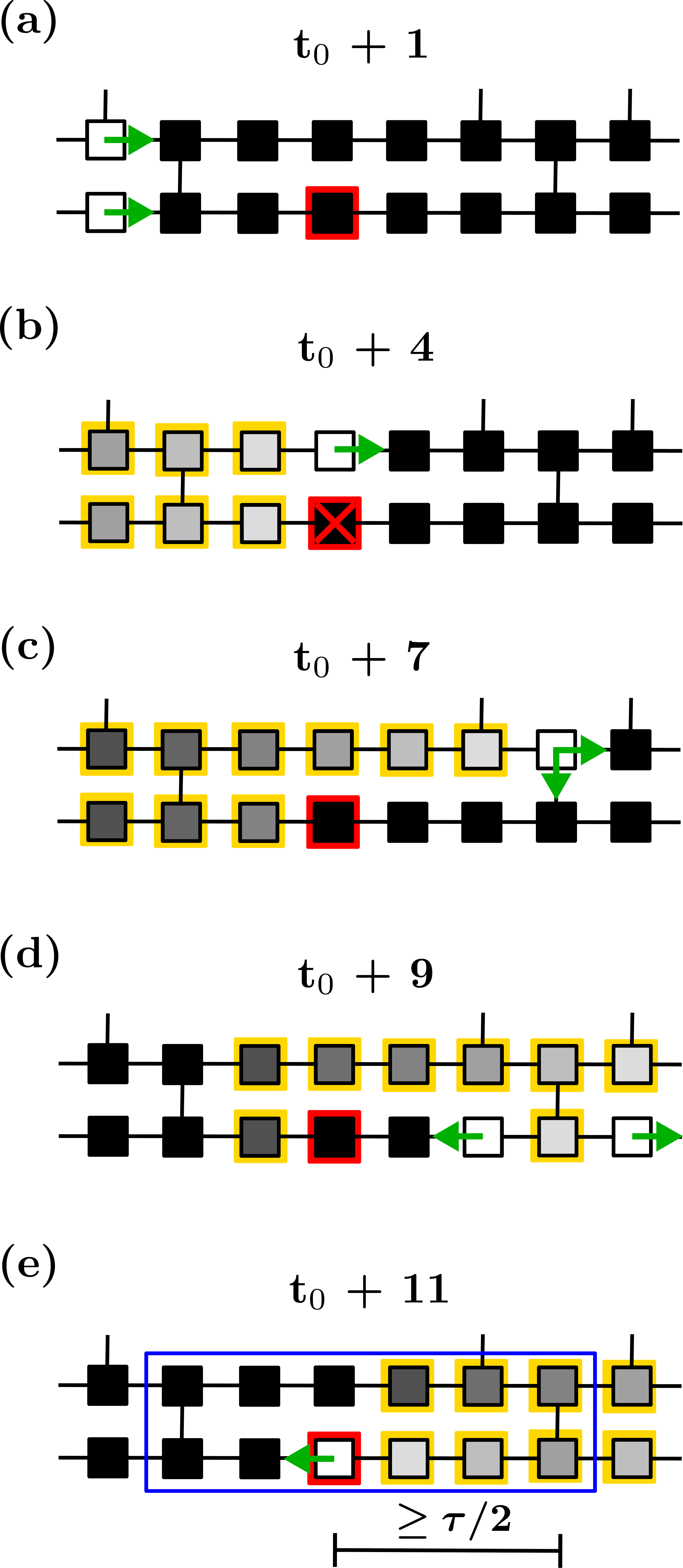}
\caption{The formation of a re-entrant circuit in the CMP model. The node that is susceptible to conduction block is marked by a red square. (a) An incoming planar wavefront (green arrows) reaches the susceptible node. (b) The node fails to excite (red cross), blocking the progression of the wavefront in the lower fibre. The wavefront advances in the upper fibre, reaching the node with a transversal connection to the lower fibre. (c) At this point, the wavefront spreads both longitudinally and transversally, initiating a retrograde propagation through the lower fibre. (d) If the path denoted by the black segment includes at least $\tau/2$ nodes, the re-entering wavefront will not encounter refractory nodes while propagating backward in the lower fibre. This establishes a structural (i.e.,  spatially stable) re-entrant circuit in the region surrounded by the blue rectangular box. When the conduction blocking node fails to fire again, the re-entrant circuit is terminated. The full evolution of this critical structure is shown in the supplementary material with an accompanying video \cite{SM}. \label{criticalstructure}}
\end{figure}

The pacemaker (sinus node) is placed on the left side of the 2D sheet and nodes lying on this edge regularly excite every $T$ time steps. The excitation propagates as a planar wavefront, mimicking  the coordinated contraction of the real atrial muscle. The parameters of the CMP model reflect clinical observations of real human atrial tissues \cite{lang2005, maceira2010, hansen2015, nakamura2011left, luke1991, verheule2003}. Clinical measurements are translated into model parameters, followed by a coarse-graining procedure leading to a square lattice of size $L = 200$ nodes, pacemaker period of $T = 220$ time steps, and refractory period of a node of $\tau = 50$ time steps. A single timestep in the model corresponds to approximately 3ms such that $T = 660$ms and $\tau = 150$ms. This refractory period is relatively short and corresponds to what may be seen clinically during burst pacing. The dynamics of the model are maintained under changes of $\tau$, but the transition from sinus rhythm to fibrillation takes place at a different point in the coupling phase space. The longer (shorter) the refractory period, $\tau$, the smaller (larger) the coupling value, $\nu_\perp$, needs to be to induce AF \cite{kishanthesis}.

The CMP model reveals that re-entrant circuits may emerge due to a combination of the electrical signal propagating on the branching structure of a heart muscle network, the three-state dynamics of nodes, and the occurrence of nodes susceptible to unidirectional conduction block. These latter nodes may fail to excite in response to an excited neighbour with small probability $\epsilon$, stopping the regular propagation of the wavefront \cite{christensen2015}. The wave of excitation proceeds forwards in the adjacent fibre until it reaches a transversal connection, leaking back through the fibre in which conduction has been previously blocked. For re-entrant circuits to emerge, the segment between the re-entry point and the node that has previously failed to excite must be long enough to prevent the backward propagating wave from being stopped by unresponsive refractory nodes. This happens when the probability of transverse connections decreases, for example, due to fibrosis. In the CMP model the formation of re-entrant circuits triggers AF. These activities survive until the circuital motion of the wavefront is annihilated by a subsequent conduction block occurring within the path of the circuit (i.e.,  self-termination) or by other waves spreading from the neighbouring regions, see Fig. \ref{criticalstructure}. For full activation maps see \cite{christensen2015}; snapshots are shown in appendix~\ref{sec:act_patterns} with accompanying videos.

Note, in the CMP model nodes are coupled with probability $\nu_\perp$ across the whole tissue. However, in the real atrium only a small patch of fibrosis may be necessary to decouple fibres and induce a re-entrant circuit. Such small patches of fibrosis may be too small to see using current MRI technologies \cite{handa2019}, inhibiting effective treatment.

%These activities, known as re-entrant circuits, are the most common driver of arrhythmias, such as atrial fibrillation (AF) \cite{smith2007}.     

\subsection{Theoretical CMP model results}

The CMP model allows us to analytically compute the risk of developing AF with respect to the fraction of transversal connections $\nu_{\perp}$, as shown in Ref.~\cite{christensen2015}. The risk is defined as the likelihood that the $L \times L$ grid has at least one region that can host a simple re-entrant circuit. The probability of having at least one transversal link on a given node is 
\begin{equation}\label{complementary}
    p_{\nu_{\perp}} = 1 - \left(1 - \nu_{\perp} \right)^2.
\end{equation}
Let $\ell$ be the distance (in number of nodes) between a node that is susceptible to conduction block and the first node to the right which has at least one transversal connection. By making use of Eq.~\eqref{complementary}, we find that the probability of $\ell$ being equal to $k$ nodes is
\begin{equation}
    \mathbb{P}\left(\ell = k \right) = (1 - p_{\nu_{\perp}})^{k}p_{\nu_{\perp}}.
\end{equation}
A given region cannot sustain a re-entrant circuit if $\ell$ is strictly smaller than $\tau/2$, see Fig.~\ref{criticalstructure}. The likelihood of this event can be calculated by summing over the probabilities of $\ell$ from $0$ to $\tau/2 - 1$, 
\begin{equation}
\begin{aligned}
    \mathbb{P}\left(\ell < \tau/2 \right) = \sum \limits_{j = 0}^{\tau/2 - 1}(1 - p_{\nu_{\perp}})^{j}p_{\nu_{\perp}}
     = 1 - ( 1 - \nu_{\perp})^\tau.
\end{aligned}
\end{equation}
Because the average number of nodes that are susceptible to conduction block is $\delta L^2$, the risk, $R$, of having at least one region that can host a re-entrant circuit is the complementary of the probability that the segments departing from these nodes are shorter than $\tau/2$, 
\begin{equation}\label{theorycurve}
\begin{aligned}
    R = 1 - \left(\mathbb{P}\left(\ell < \tau/2 \right)\right)^{\delta L^2}
 %                   = 1 - \left( \sum \limits_{j = 0}^{\tau/2 - 1}(1 - p_{\nu_{\perp}})^{j}p_{\nu_{\perp}} \right)^{\delta L^2}\\
                    = 1 - \left[1 - ( 1 - \nu_{\perp})^\tau \right]^{\delta L^2}.
\end{aligned}
\end{equation}
Equations (1)-(4) have been derived in \cite{christensen2015}. Equation ~\eqref{theorycurve} provides a simple analytical tool to estimate the risk of developing AF in the CMP model. The result indicates that the risk of AF increases as the tissue becomes more decoupled/fibrotic, in agreement with the current clinical understanding \cite{chelu2018}. Likewise, the theory predicts that the risk of fibrillation increases as the size of the atrial tissue increases, in agreement with clinical practice where left atrial volume is used as a predictor of the risk of developing AF \cite{tsang2001}. The theoretical analysis presented here has additional value in that we can predict how the model will change if the rules or parameters are changed, allowing for a comparison with similar computational models of AF. This is discussed in detail in section~\ref{CMP_Context}.

\begin{figure}
\centering
\includegraphics[width=8.0cm, height = 7.5cm]{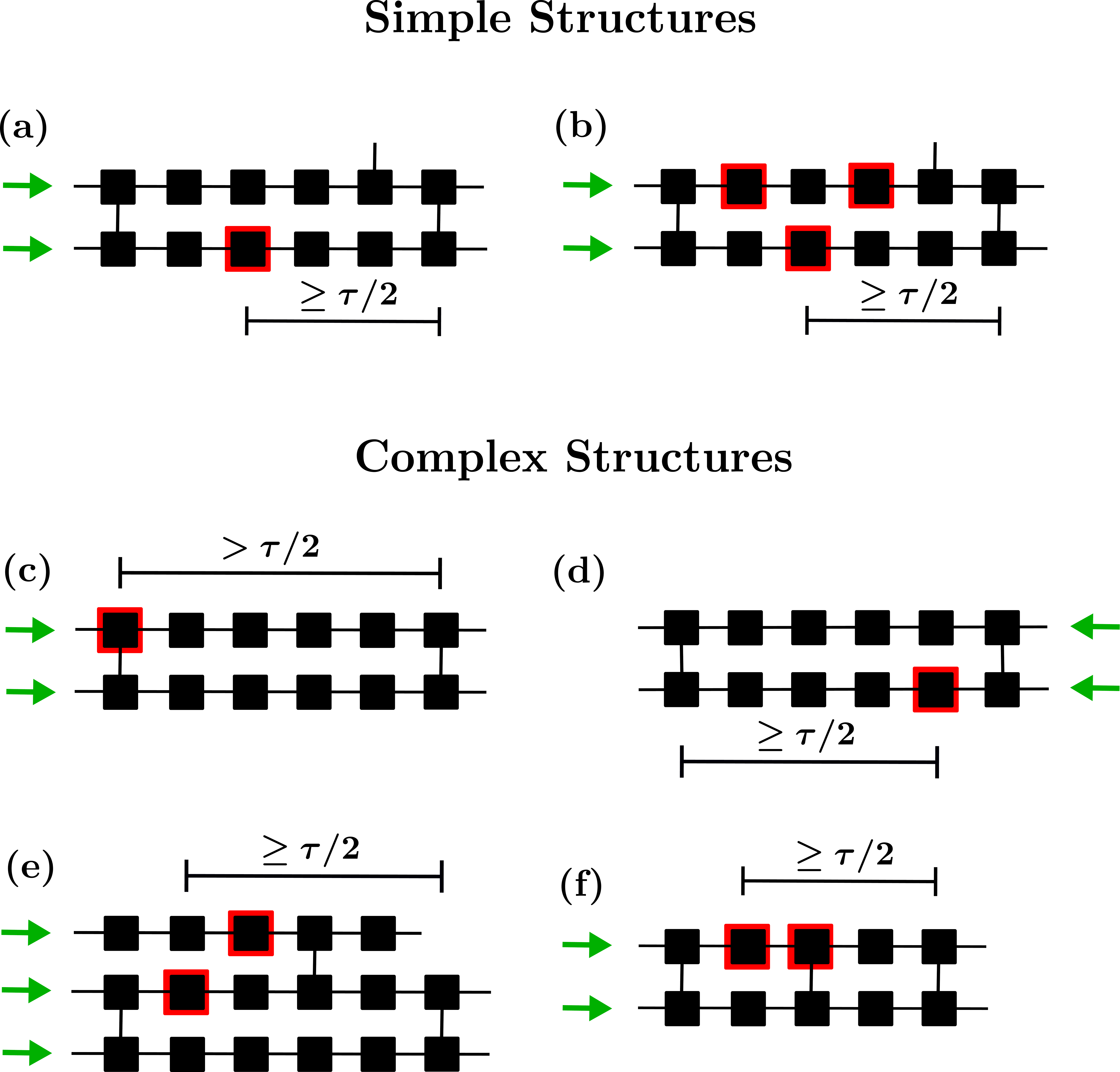}
\caption{
Critical structures in the CMP model. The black segment on top of each structure represents the minimum distance (in number of nodes) between the relevant conduction blocking node (red squared border) and the first regular node to the right which has at least one transversal connection for the structure to sustain a re-entrant circuit. The wavefront direction is indicated by the green arrows. (a)-(b) Simple critical structures are triggered by a single block of the incoming planar wavefront originating from sinus rhythm. These structures might include multiple nodes that are susceptible to conduction block, increasing the probability of self-termination. (c)-(f) The activation of complex critical structures requires a sequence of conduction blocks of the planar wave front or waves of excitation not originating from sinus rhythm. The probability of triggering these regions is much smaller than in (a)-(b). (c) The presence of at least one transversal connection departing from the conduction blocking node makes the activation more difficult as this node must fail to excite twice before prompting a re-entrant circuit. (d) This structure cannot be triggered from sinus rhythm but it can be triggered by a single block of a wave of excitation originating from elsewhere. (e)-(f) The activation of these structures requires multiple blocks of the planar wavefront to occur in different nodes. Examples of the evolution of each structure are shown in the supplementary material with accompanying videos \cite{SM}.
\label{structuretypes}}
\end{figure}

This theoretical result builds on the assumption that re-entrant circuits form from the failure of a single conduction blocking node. However, this assumption does not account for all instances in which AF is triggered in the model. For instance, the probability of triggering a re-entrant circuit varies across the lattice depending on the architecture of the hosting region, see Fig.~\ref{structuretypes}. Notably, some re-entrant circuits may only activate if two nodes susceptible to conduction block fail successively (i.e., in a single activation cycle).

These details indicate that the CMP theory represents an ideal case for AF driven by simple re-entrant circuits only. The theory assumes that if a simple circuit exists, the tissue spends 100\% of the time in AF. Therefore, the theory curve sets a limit on the maximum time the model can spend in AF due to simple circuits only.

\subsection{Model Behavior \label{MB}}
Local regions that are capable of hosting re-entrant circuits are called critical structures, see Fig.~\ref{criticalstructure}. A critical structure is active (inactive) when it hosts (does not host) a re-entrant circuit. In the CMP model, critical structures are classified according to the complexity of their activation and deactivation mechanisms. Structures which can activate and terminate from the failure of a single conduction blocking node from sinus rhythm are referred to as simple. This includes cases where a critical structure contains multiple conduction blocking nodes, but only one must fail to allow for the formation of a re-entrant circuit. All other configurations in which the planar wavefront from sinus rhythm requires multiple conduction blocks to fail to form a re-entrant circuit are referred to as complex. The latter class includes critical structures that are only triggered by waves of excitation not originating from sinus rhythm (proceeding from right to left on the lattice), see Fig~\ref{structuretypes}(d).

For large values of $\nu_{\perp}$, the model is in sinus rhythm indefinitely. The high number of transversal connections excludes the presence of regions that are critical for AF initiation and preservation as there are no sections of length $\geq \tau/2$ without a transverse connection. When $\nu_{\perp}$ decreases, for example due to increasing fibrosis \cite{spach1997}, we observe a more pronounced branching structure of the lattice which favours the spontaneous emergence of structures that can host re-entrant circuits. This increases the risk of developing AF.

When $\nu_{\perp}$ is sufficiently small, increasing $\delta$ extends the time the system spends in AF. This occurs because a larger fraction of nodes are susceptible to conduction block and this increases the number of regions that can host a re-entrant circuit. However, the sensitivity of the system to the fraction of conduction blocking nodes, $\delta$, rapidly vanishes as $\nu_{\perp}$ increases, suggesting that weak branching prevents the formation of critical structures independent of the fraction of nodes that are susceptible to conduction block \cite{kishanthesis}. The probability that a conduction blocking node fails to excite, $\epsilon$, does not significantly influence the relationships between $\nu_{\perp}$ and the fraction of time the system spends in AF \cite{kishanthesis}. This implies that $\epsilon$ is mainly used to set the time scale of the model. More precisely, for simple re-entrant circuits, we note that $\epsilon$ does not appear in the derivation of the risk of AF in Eq.~\eqref{theorycurve}. This is because $\epsilon$ effects both the probability that a simple re-entrant circuit activates and deactivates. If $\epsilon$ is reduced, it will, on average, take longer for a simple re-entrant circuit to activate. However, once active, that re-entrant circuit will take longer to de-activate than the equivalent circuit with a larger value of $\epsilon$. That means that $\epsilon$ determines the duration of paroxysmal AF episodes and the time between paroxysmal AF episodes, but has a minimal effect on the overall risk of AF in the CMP model. Likewise, $\epsilon$ has no effect on the period of any simple re-entrant circuits formed. However, if circuits exist with an asymmetry between the probability of activation and deactication, $\epsilon$ may play a role in the duration of individual fibrillatory events.

The length of the refractory period, $\tau$, sets the minimum distance between the conduction blocking node and the first regular node to the right which has at least one transversal connection for the structure to sustain a re-entrant circuit, see Fig.~\ref{criticalstructure} and Fig.~\ref{structuretypes}. Given a fixed value of $\delta$, lowering $\tau$ increases the number of regions that can host re-entrant circuits, increasing the time the system spends in AF.

In the CMP model, the system is defined to have entered AF if the number of active nodes per time step $a(t)$ exceeds $1.1 \times L$ (220) nodes for $T$ consecutive time steps,
\begin{equation}\label{pinducing}
    p^{\text{AF}}_{\text{CMP}}(t) = \begin{cases} 1 &\quad \text{if min$([a(t-T),\ldots,a(t)]) \geq 220$} \\
    0 &\quad \text{otherwise}
    \end{cases}
\end{equation}
where $t$ can take integer values in the range $t = T, \dots, S$ and $S$ is the duration of the experiment (in time steps) \footnote{Note, it is coincidental in this case that $T = 220 = 1.1 L$}.
We use Eq.~\eqref{pinducing} to study how the probability of inducing AF varies with the amount of coupling $\nu_{\perp}$ and compare this statistic with its theoretical estimations, see Eq.~\eqref{theorycurve}. Note, Eq.~\eqref{pinducing} gives a working definition of AF in the CMP model and was derived by inspection in previous work \cite{christensen2015,manani2016,kishanthesis}. 

The definition used here is not unique and is not robust against changes in the pacing frequency $T$. The definition is designed to measure whether nodes in the model are activated more frequently than would be expected in sinus rhythm. This is based on the principle that if nodes are being activated at a rate higher than the pacing rate, then there must a source of fibrillatory wavefronts other than the sinus node. 
A superior method would be to measure the average activation frequency of nodes relative to the pacing frequency explicitly, rather than the number of active nodes, since this would be more robust against changes in $T$. 
However, to be consistent with previous work we use the existing definition in the current paper. We stress that for fixed $T$, the two methods give almost identical results. Both methods compare well with a clinical definition of AF where AF is diagnosed from ECG or electrogram recordings, see appendix~\ref{sec:ECG}. We do not generate electrograms as standard in the CMP model since this significantly increases the computational burden of the simulations. Additionally, we do not explicitly distinguish between AF and atrial tachycardia (AT) in the CMP model; see appendix~\ref{sec:MF_vs_cCMP} for further details. 
%although in practice AT can be thought of as high frequency activity driven by a single re-entrant circuit whereas AF refers to two or more competing re-entrant circuits.

\begin{figure}
\centering
\includegraphics[width=7.0cm]{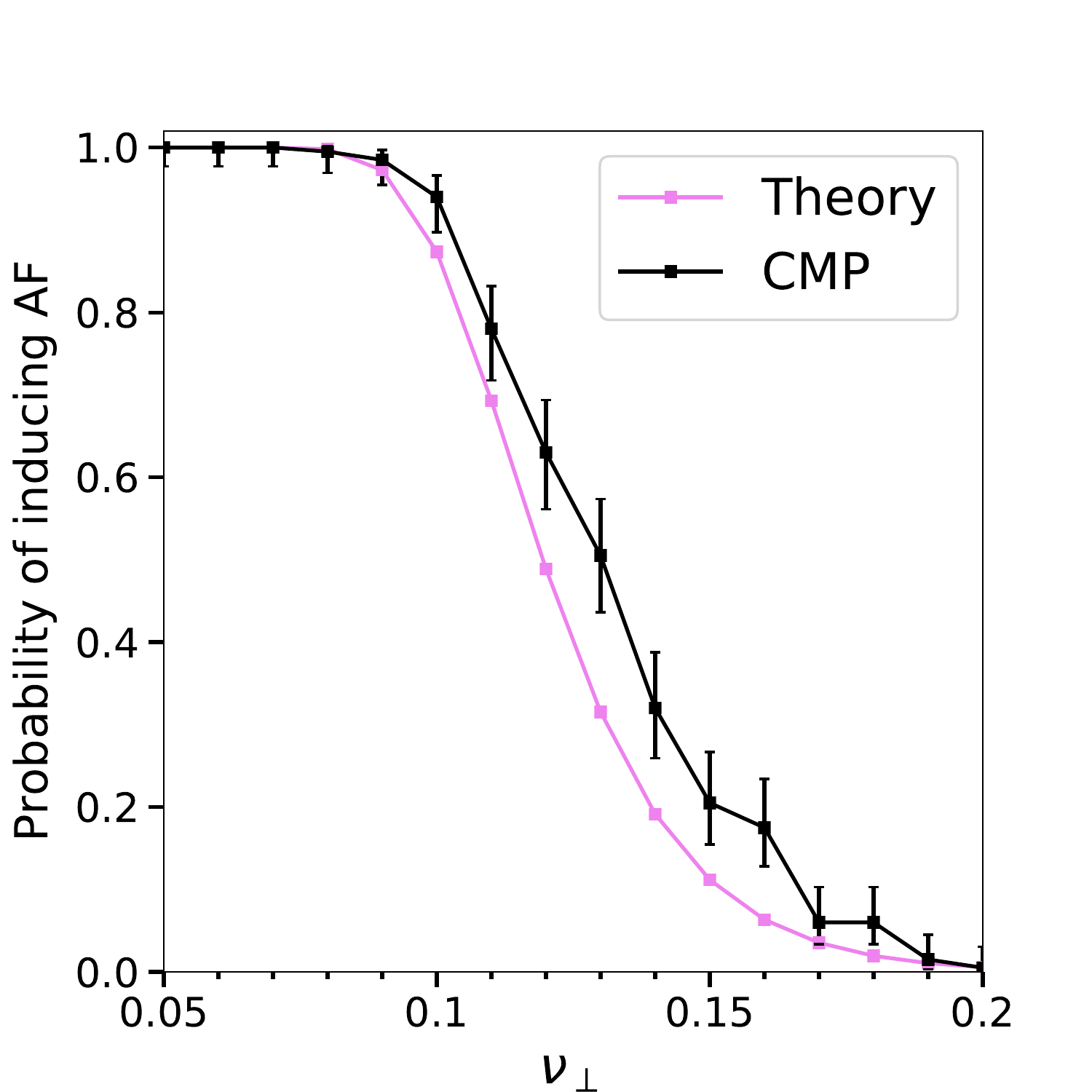}
\caption{Phase diagram of the probability of inducing AF as a function of the fraction of transversal connections $\nu_{\perp}$. 
%The settings of the model are: $L = 200$, $T = 220$, $\tau = 50$, $\delta = 0.01$ and $\epsilon = 0.05$. 
The violet line represents the theoretical risk curve, see Eq.~\eqref{theorycurve}. For each value of $\nu_{\perp}$, we perform 200 simulations of the CMP model and compute the average probability of inducing AF (black square), see Eq.~\eqref{pinducing}.
%Error bars show the Agresti-Coull 97.5\% binomial confidence intervals of the sample average. 
The duration of each simulation is $S = 10^6$ time steps. 
For both the model and the theory, we observe that the system never (always) develops AF for $\nu_{\perp} \gtrapprox 0.2$ ($\nu_{\perp} \lessapprox 0.1$). Within this interval, the probability of developing AF rapidly increases as $\nu_{\perp}$ is lowered. For any value of $\nu_{\perp}$ between 0.1 and 0.2, the probability of inducing AF in the CMP model (black) is always higher than in the CMP theory (violet).\label{phasediagram}}
\end{figure}

\begin{figure*}
\centering
\includegraphics[width=\linewidth]{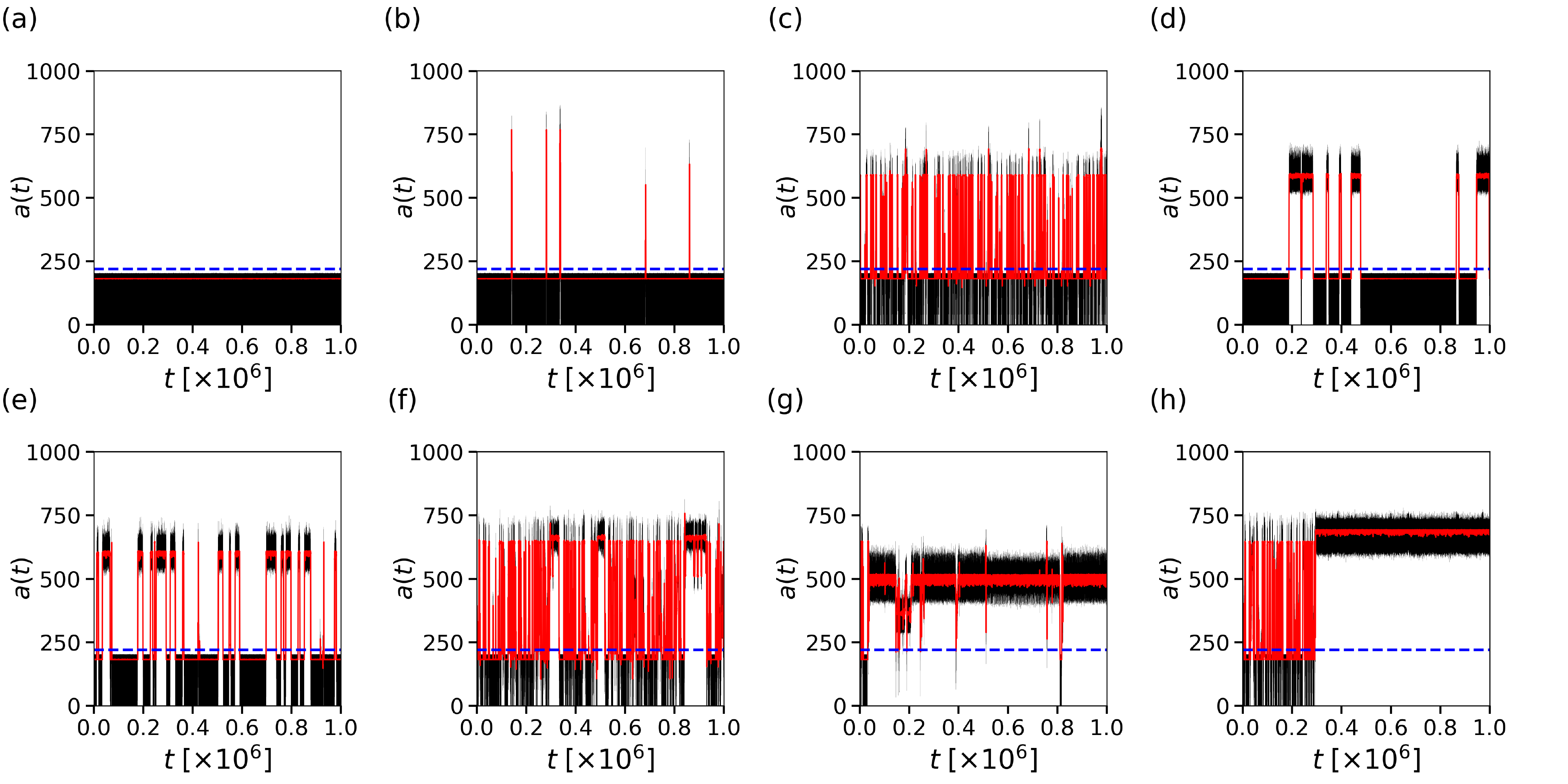}
\caption{
The number of excited nodes per time step $a(t)$ (black thin line) and its moving average $\langle a(t) \rangle$ calculated over $T = 220$ successive time steps (red solid line) in eight different simulations of the CMP model. All simulations are generated with identical model parameters. The coupling value is set at $\nu_{\perp} = 0.11$.
The system is in AF when the number of excited nodes per time step exceed 220 (blue dashed line) for at least $T$ time steps. 
The figure demonstrates the broad spectrum of AF persistence that naturally emerges in the CMP model, from (a) sinus rhythm, through (b)-(g) various forms of paroxysmal AF, to (h) fully persistent AF. The figures exhibit a range of different event times, and asymmetries between the period of time in and out of AF. Subfigures (b), (e) and (h) are dominated by short, intermediate, and long AF events respectively. Subfigures (f) and (g) exhibit an interplay between short and intermediate, and intermediate and long event times respectively. These figures demonstrate that complex behaviour can emerge at the model macrostructure from specific details at the model microstructure, independent of the parameters of the model.
\label{modeldynamics}}
\end{figure*}

The probability of inducing AF in simulations of the CMP model is systematically higher than in the CMP theory, see Fig.~\ref{phasediagram}. These findings are somewhat surprising since the CMP theory assumes the most favourable conditions for the emergence of AF from simple re-entrant circuits only. We assert that this excess could be explained by the fact that re-entrant circuits in the CMP model might have multiple mechanistic origins that are not accounted for in the CMP theory. Furthermore, the CMP theory assumes that re-entrant circuits are triggered by single unidirectional conduction blocks, that is, AF is exclusively driven by simple critical structures, see Fig.~\ref{structuretypes}.

To better understand the discrepancy between theory and experiment, we look at the trace of the number of active nodes in the model. AF is paroxysmal when this statistic exhibits large fluctuations which prevent it from stabilizing above the AF threshold, i.e., the number of active nodes frequently falls below 220 nodes with only short periods of high frequency activity. AF is persistent when the number of active nodes consistently exceeds the AF threshold for extended periods of time. If AF in the model has a unique mechanistic origin, we would expect tissues at the same level of coupling to exhibit statistically similar behaviors in the number of the active nodes over time. However, we find that this is not the case -- there is significant heterogeneity among systems characterized by the same parameters, e.g.,  the amount of uncoupling, or the fraction of conduction blocking nodes.

In Fig.~\ref{modeldynamics} all tissues are generated using the same parameters, with $\nu_{\perp} = 0.11$. Tissue (a) remains in sinus rhythm indefinitely. Tissue (b) remains mostly in sinus rhythm, with rare fibrillatory events on the order of $10^3$ time steps in the model. In real time, these events are on the order of $1s$. It is plausible that clinically, such short events may be interpreted as an ectopic beat rather than AF. From tissue (c), through to tissue (g), we observe a spectrum of AF persistence. This includes short frequent events in (c), rare intermediate events in (d), frequent intermediate events in (e), a combination of short and intermediate events in (f), and long events with brief interruptions in (g). Only in (h) do we see a permanent transition from short paroxysmal AF, to persistent AF. The event shown in (h) is on the order of $30$ minutes when converting to real time. Repeating those simulations where persistent AF appears to last until the end of the simulation, these simulations are extended to $10^9$ time steps without the simulation reverting to sinus rhythm. In real time, these events are on the order of $1$ month. For practical reasons, we have not investigated events on timescales longer than $10^9$ time steps. Note, that for visual clarity, the example chosen in Fig.~\ref{modeldynamics}(h) is driven by a single dominant driver which may be defined as AT rather than AF. However, in most cases, persistent activity is maintained for long time periods with the presence of multiple competing drivers, see appendix~\ref{sec:act_patterns} for an example. 

The variability in the persistence of AF in the CMP model has been studied previously in \cite{manani2016}. The authors focused on the relationship between the amount of uncoupling in the lattice (i.e.,  $\nu_{\perp}$) and the features of the developed AF in 32 independent experiments. In agreement with clinical observations \cite{kottkamp2013, veasey2015}, they report high degrees of heterogeneity in the progression to persistent AF and in the amount of uncoupling required for AF to emerge. Similarly to Fig.~\ref{modeldynamics}, they observe very different AF patterns across systems characterized by the same amount of uncoupling, asserting that the emergence of re-entrant circuits is subject to the local distribution of transversal connections, not the global amount of coupling, i.e.,  $\nu_{\perp}$. However, the authors do not satisfactorily explain how and why different AF patterns emerge from the microstructure of the CMP model.

The findings presented in Fig.~\ref{phasediagram} and Fig.~\ref{modeldynamics} provide two important pieces of evidence against the assumption that AF is exclusively driven by simple re-entrant circuits. First, they show that the probability of inducing AF is systematically higher in the CMP model than in the CMP theory, see Fig.~\ref{phasediagram}. Second, they reveal different activation patterns do not appear consistent with simple structures activating and deactivating with fixed rates. Individual events exhibit a spectrum of lifetimes before reverting to sinus rhythm, from seconds to months. This motivates us to assess whether different mechanistic origins of AF are effectively present in the CMP model and how they eventually relate with the progression to persistent AF from paroxysmal AF.

In the following sections, we take up these challenges by removing layers of complexity from the CMP model. This allows us to derive simpler frameworks in which we can examine whether re-entrant circuits have different mechanistic origins and how the features of these activation processes influence the development of AF. 

In Section \ref{MF}, we start with the simplest approach by removing all the spatial elements of the CMP model. This is done by condensing the CMP model into a mean-field (MF) model in which complex critical structures and interactions between re-entrant circuits (i.e.,  wave collisions) are neglected. This simple framework allows us to study AF under the assumption that fibrillation is exclusively driven by independently activated simple re-entrant circuits. We show that the MF model systematically underestimates the probability of inducing AF and the persistence of AF. 

In Section \ref{FUNCTIONAL} we dissect this discrepancy by reintroducing the spatial elements of the CMP model while carefully controlling the placement of nodes susceptible to conduction block. This prevents the formation of complex critical structures. The main advantage of this controlled CMP model (cCMP) over the simpler MF model is that it allows us to quantify how different activation mechanisms contribute to AF emergence and maintenance. Like the MF model, the cCMP model underestimates the probability of inducing AF and the persistence of AF with respect to the CMP model. However, the cCMP model does not increase the time in AF relative to what is found in the MF model with the exception of very small fluctuations explained by differences in individual event times. 

Finally, we confirm that the difference in the probability of inducing AF and the persistence of AF between the CMP and cCMP models stems from the contribution of complex re-entrant circuits which exhibit an asymmetry between the probability of activating and deactivating a re-entrant circuit. These complex structures may only require a single failure from a conduction blocking node to initiate, but multiple failures to terminate, resulting in long individual event times. Additionally, these structures may be coupled as part of a larger critical structure such that the termination of a re-entrant circuit anchored to a specific sub-structure immediately initiates a new re-entrant circuit in a coupled sub-structure. We demonstrate these mechanisms explicitly and show that as the probability that a node is susceptible to conduction block is lowered, the spatial density of conduction blocking nodes falls to the extent that multiple failing nodes are not required for the termination of a re-entrant circuit. As a result, the time the CMP and cCMP models spend in AF collapse onto a single curve. This demonstrates that an increase in the local density of conduction blocking nodes is highly proarrhythmic.

\section{\label{MF} Mean-Field Model of AF}
In the CMP model, critical structures activate and deactivate to sustain AF. Initially, the system is in sinus rhythm as planar waves of excitation released from the sinus node (pacemaker) propagate on the lattice. The motion of the planar waves is disrupted now and then by conduction blocks occurring across the grid. At some point in time, a conduction block forms the initial re-entrant circuit. This re-entrant circuit cannot maintain AF indefinitely because it will either self-terminate or be terminated by waves spreading from the surrounding regions. However, its circuital motion intensifies the model activity, generating disorganized, high-frequency activation wavefronts that spread across the lattice. When the system enters this state, non-planar waves of excitation spreading from the active re-entrant circuit reach dormant critical structures at a much higher frequency than the pacemaker waves. This initiates a chain of asynchronous activations and deactivations of different critical structures located across the lattice, protracting the current AF episode until the complete disappearance of re-entrant circuits brings the system back to sinus rhythm.

In the CMP model, it is unclear whether these interactions between simple critical structures are the only drivers of AF. In particular, the results discussed in Fig.~\ref{phasediagram} and Fig.~\ref{modeldynamics} motivate us to examine whether other activation mechanisms drive AF and how differences between paroxysmal and persistent AF emerge. The simplest approach to this problem is to derive a framework in which fibrillation is solely driven by independently activated simple re-entrant circuits and to compare AF-related statistics against the CMP model. 

\begin{figure}
\centering
\includegraphics[width=7.0cm]{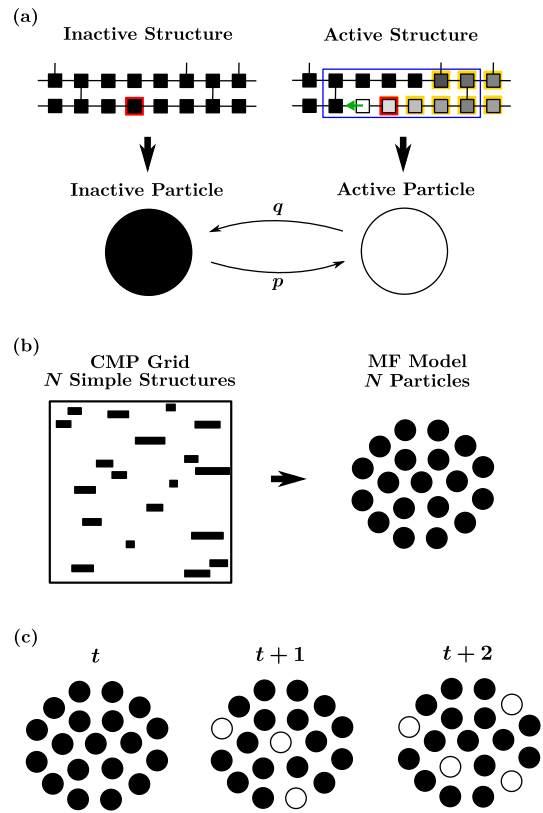}
\caption{The CMP model is condensed into a mean-field (MF) model of AF. (a) Simple critical structures are mapped into particles which can take two distinct states: active (i.e.,  hosting a re-entrant circuit, blue path) or inactive. We enforce the following assumptions: 
i) the location (spatial positioning) of a particle is irrelevant,
ii) all particles activate with rates $\epsilon/T$ when the system is in sinus rhythm and $\epsilon/\langle \ell \rangle$ when the system is in AF,
iii) all particles deactivate with rate $\epsilon/\langle \ell \rangle$,
iv) all particles have the same length $\langle \ell \rangle$  and
v) particles can change their states at any time step. (b) Simple critical structures (black filled rectangles) found in the CMP lattice are condensed into particles (black filled circles). (c) The evolution of the MF system is driven by $N$ independent particles that activate and deactivate with probability $p$, and $q$ respectively, depending on the current state of the particle and the system.}
\label{meanfieldscheme}
\end{figure}

To do so, we translate the features of the CMP lattice into a simple mean-field (MF) model of AF in which $N$ particles independently turn on and off. For a one-to-one comparison, the number of particles, $N$, is directly observed from the number of simple critical structures present in the CMP lattice at a given level of coupling. The fact that simple critical structures are characterized by a few well defined architectural features allows us to systematically inspect the grid and detect each region falling into this category. 

In the MF model, the system is represented by a simple Markov chain. At a given time $t$, the state of the chain is the number of active particles $N_{\text{a}}(t)$, such that $t:N_{\text{a}}(t) = \{0,1,\dots,N\}$. When $N_{\text{a}}(t) = 0$, the system is defined as being in sinus rhythm where any existing critical structure has a chance to be triggered every $T$ time steps (pacemaker frequency). On the other hand, $N_{\text{a}}(t) \geq 1$ is defined as the MF model exhibiting AF. In this case, the length of active re-entrant circuits sets the frequency (in time steps) at which inactive critical structures can be triggered. For the sake of simplicity, we assume that particles have the same length $\langle \ell \rangle$ corresponding to the average length (in number of nodes) of the simple critical structures tracked across the CMP lattice. At any time step, inactive particles activate with rate $p$ and active particles deactivate with rate $q$, see Fig.~\ref{meanfieldscheme}.

Activation rates change depending on the state of the system, mimicking the fact that the presence of at least one re-entrant circuit significantly increases the frequency at which dormant critical structures can be triggered. It follows that $p$ and $q$ are given by:
\begin{subequations}
\begin{align} \label{rates_a}
p &= \begin{cases} \frac{\epsilon}{T}, & \mbox{if } N_a = 0, \\ \frac{\epsilon}{\langle \ell \rangle}, & \mbox{if } N_a > 0, \end{cases}\\
\label{rates_b}
q &= \frac{\epsilon}{\langle \ell \rangle}.
\end{align}
\end{subequations}

The probability $P_{i,j}$ of transitioning from $i$ to $j$ active particles is analytically derived,
\begin{subequations}
\begin{align}
\begin{split}\label{firstp}
P_{i,j} &= \smashoperator{\sum_{k=0}^{\min\{i, N-i\}}}\text{B}(i,k,q)\text{B}(N-i,j-i+k,p) \;\; \;\;\mbox{if } j \geq i
\end{split}\\
\begin{split}\label{secondp}
P_{i,j} &= \smashoperator{\sum_{k=i-j}^{\min\{i, N-i\}}}\text{B}(i,k,q)\text{B}(N-i,j-i+k,p) \;\;\;\;  \mbox{if } j < i
\end{split}
\end{align}
\end{subequations}
where $\text{B}(N,k,r)=\begin{pmatrix} N \\ k \end{pmatrix} r^k (1-r)^{N-k}$ is the binomial distribution yielding the probability of getting exactly $k$ successes in $N$ trials when the
probability of success is $r$ \cite{kishanthesis}.

This simple model allows us to calculate the same AF-related statistics that one can compute in simulations of the CMP model. For instance, we can easily adapt Eq.~\eqref{pinducing} to the MF model
%\begin{equation}\label{pinducing2}
%    p^{\text{AF}}_{\text{MF}} = \mathbbm{1}_{\{N_{a}(t) \geq 1\}}
%\end{equation}
\begin{equation}\label{pinducing2}
    p^{\text{AF}}_{\text{MF}}(t) = \begin{cases} 1 &\quad \text{if $N_a(t) \geq 1$,} \\
    0 &\quad \text{otherwise,}
    \end{cases}
\end{equation}
where $t = 1, \dots, S$ and $S$ is the duration of the experiment (in time steps). In addition, we are interested in measuring the time the system spends in AF as a function of the amount of coupling $\nu_{\perp}$. In the MF model, this statistic corresponds to the ratio between the number of time steps in which at least one particle is active and the duration of the experiment 
\begin{equation}\label{TAF}
    T^{\text{AF}}_{\text{MF}} = S^{-1}\sum\limits_{t = 1}^{S}p^{\text{AF}}_{\text{MF}}(t).
\end{equation}

By taking a continuous approximation and deriving a master equation, a full analytic solution can be derived for the fraction of time the model spends in AF, 
\begin{equation}
T^{\text{AF}}_{\text{cMF}} = 1-\tilde{p}(0) = \frac{2^N-1}{2^N +p/p_0 -1},
\label{cMFequation}
\end{equation}
where $p_0$ is the probability of activating a particle when $N_{a} = 0$, and $\tilde{p}(0)$ is the probability that no particles are active, see appendix~\ref{CMF} for the derivation. Equation~\eqref{cMFequation} indicates that the fraction of time spent in fibrillation approaches 1 only when $N \rightarrow \infty$. However, for finite $N$, the fraction of time in AF is finite. In the case of $N=2$, $T^{\text{AF}}_{\text{cMF}} =0.405$, whereas for $N=10$, $T^{\text{AF}}_{\text{cMF}} = 0.996$. While the $N=10$ case may explain very long individual events, for the examples shown in Fig.~\ref{modeldynamics} at $\nu_{\perp}=0.11$, the average value of $N$ is approximately $3$, see appendix~\ref{sec:MF_vs_cCMP}, giving $T^{\text{AF}}_{\text{cMF}} = 0.614$. Hence, the persistent AF observed in Fig.~\ref{modeldynamics}(h) is not explained by the simple birth-death like dynamics of simple re-entrant circuits underlying the continuous and discrete MF models.

For convenience, given the discrete nature of fibrillatory events and our interest in the probability of inducing fibrillation within a given time-frame, we choose to use the discrete version of the MF model as described above for the remainder of the paper.

In the MF model, the spatial elements of the CMP model are neglected to prevent the potential formation of re-entrant circuits from collisions between multiple waves of excitation. Furthermore, the correspondence between the numbers of simple critical structures and system particles excludes any eventual contribution from complex critical structures. The number of tracked simple critical structures reflects, to a good extent, the key architectural properties of the CMP lattice, namely, the amount of coupling, i.e.  $\nu_{\perp}$, and the fraction of nodes that are susceptible to conduction block, i.e.  $\delta$. Pegging the number of particles to the number of simple critical structures allows us to calibrate the MF model with the CMP model.

\begin{figure}
\centering
\includegraphics[width=7.0cm]{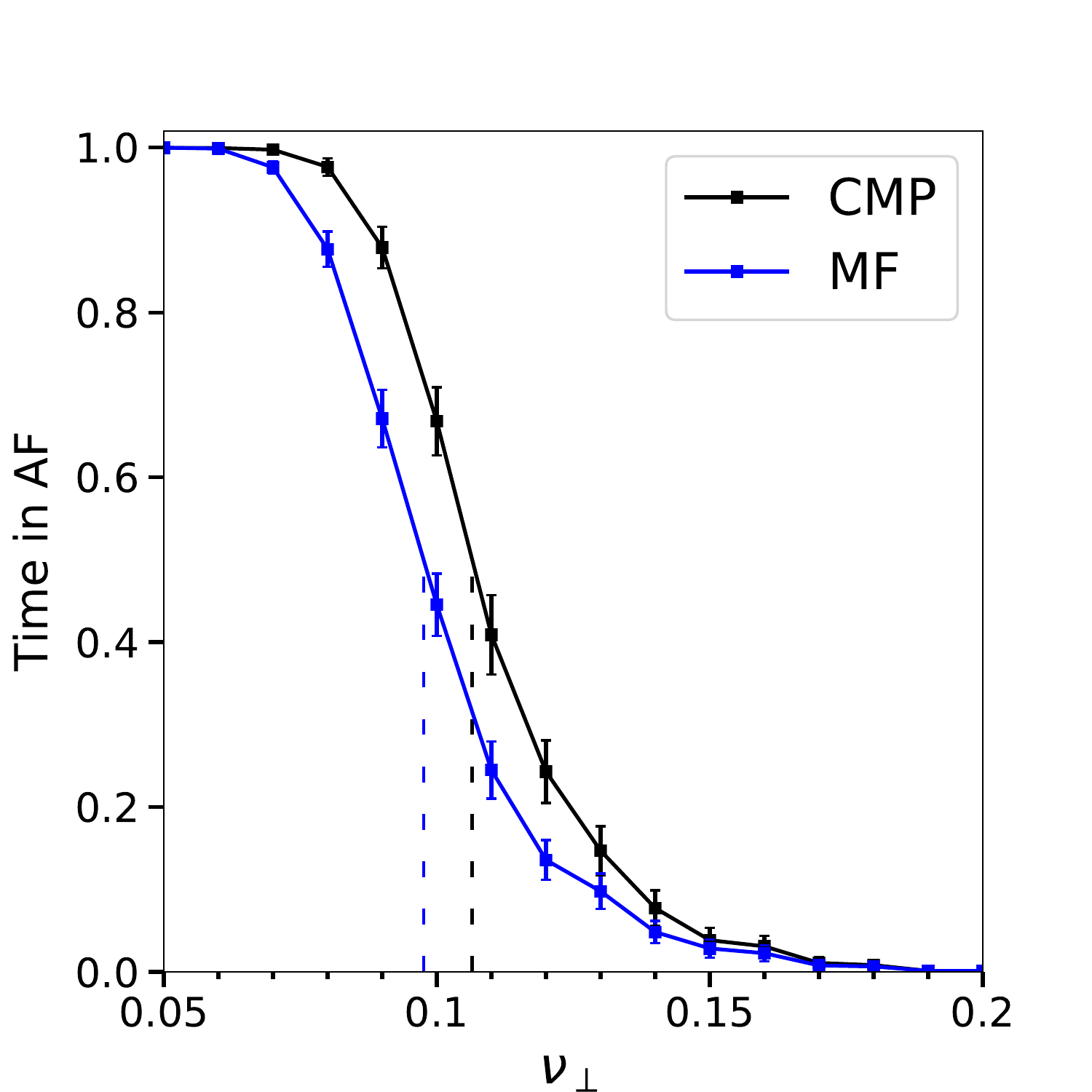}
\caption{Phase diagram of the time in AF as a function of the fraction of transversal connections, $\nu_{\perp}$, for the CMP (black) and MF (blue) models. For each value of $\nu_{\perp}$ we perform 200 simulations of the CMP model and measure the time in AF over $S = 10^6$ time steps, see Eq.~\eqref{TAFCMP}. For each simulations of the CMP model we 
derive the associated MF model, see Fig.~\ref{meanfieldscheme}, and compute the time in AF according to Eq.~\eqref{TAF}. The time the system spends in AF is significantly higher in the CMP model (black) than in the MF model (blue). Sharp transitions in the time in AF occur around the threshold values $\nu^{*}_{\perp} \approx 0.11$ (CMP, black dashed line) and $\nu^{*}_{\perp} \approx 0.09$ (MF, blue dashed line). \label{mfphasediagram}}
\end{figure}

To compare the MF model with the CMP model, we define the time in AF for the CMP model equivalently to Eq.~\eqref{TAF},
\begin{equation}\label{TAFCMP}
    T^{\text{AF}}_{\text{CMP}} = S^{-1}\sum\limits_{t = 1}^{S}p^{\text{AF}}_{\text{CMP}}(t),
\end{equation}
where $a(t)$ is the number of active nodes at time step $t$. The MF model spends significantly less time in AF than the CMP model, see Fig.~\ref{mfphasediagram}. This may be because the neglected spatial features of critical structures, such as different lengths and asynchronous activation and deactivation, have a significant role in AF emergence and maintenance. However, an enhanced version of the MF model, see appendix~\ref{extendedMF}, reintroduces these omissions and indicates that these changes have no noticeable effect of the time the MF model spends in AF. Alternatively, higher order structures may exist which provide an additional contribution to the time in AF in the CMP model.
%On the other hand, AF in the CMP model might be driven by mechanisms other than simple re-entrant circuits, such as multiple conduction blocks, or critical circuits that cannot be initiated from sinus rhythm. 
%In appendix \ref{scaledMF}, we consider the case of the MF model with twice the number of critical structures to simulate the effect that non-sinus rhythm re-entrant circuits might have on the time in AF. However, although these additional particles increase the time the MF model spends in AF, the total time in AF still falls below the CMP model. Hence, in the following sections we investigate the effect of temporally stable re-entrant circuits formed from wavefront collisions, and the effect multiple conduction blocks.

\section{\label{FUNCTIONAL} Emergence of Complex Critical Structures}

\subsection{The Controlled CMP Model}

The problem of detecting the variety of activation mechanisms and disentangling their roles in the patterns of AF can be approached in two ways. We may attempt to devise a detection algorithm to identify all simple and complex critical structures located across the lattice and assess their contribution to the phase diagrams in Fig.~\ref{mfphasediagram}. However, the wide variety and complexity of these structures poses a significant challenge with no easy method to verify that all circuits have been detected.

A more straightforward approach involves deliberately constructing simple re-entrant circuits by controlling the placement of conduction blocking nodes across the lattice, referred to as the controlled CMP model (cCMP). For simplicity, we achieve this by identifying the isolated segments of length $\geq\tau/2$ in the lattice and note which nodes in the segment, if susceptible to conduction block, would form a simple re-entrant circuit consistent with those simple structures shown in Fig.~\ref{structuretypes}. A fraction $\delta$ of these nodes are set to be susceptible to conduction block. All other nodes are not susceptible to conduction block. This leads to a special CMP lattice in which AF is driven by simple re-entrant circuits only since conduction blocking nodes are only found in simple critical structures and not across the lattice as a whole.

To compare the cCMP and CMP models, we make a copy of the cCMP lattice and randomly place conduction blocking nodes across the lattice as a whole with probability $\delta$ -- this model is equivalent to the regular CMP model. We simulate the two models and compare the probability of inducing AF and the time in AF. In this scenario, the eventual differences in AF related statistics quantify the contribution of complex critical structures to AF persistence and maintenance.

\begin{figure}
\centering
\includegraphics[scale=0.4]{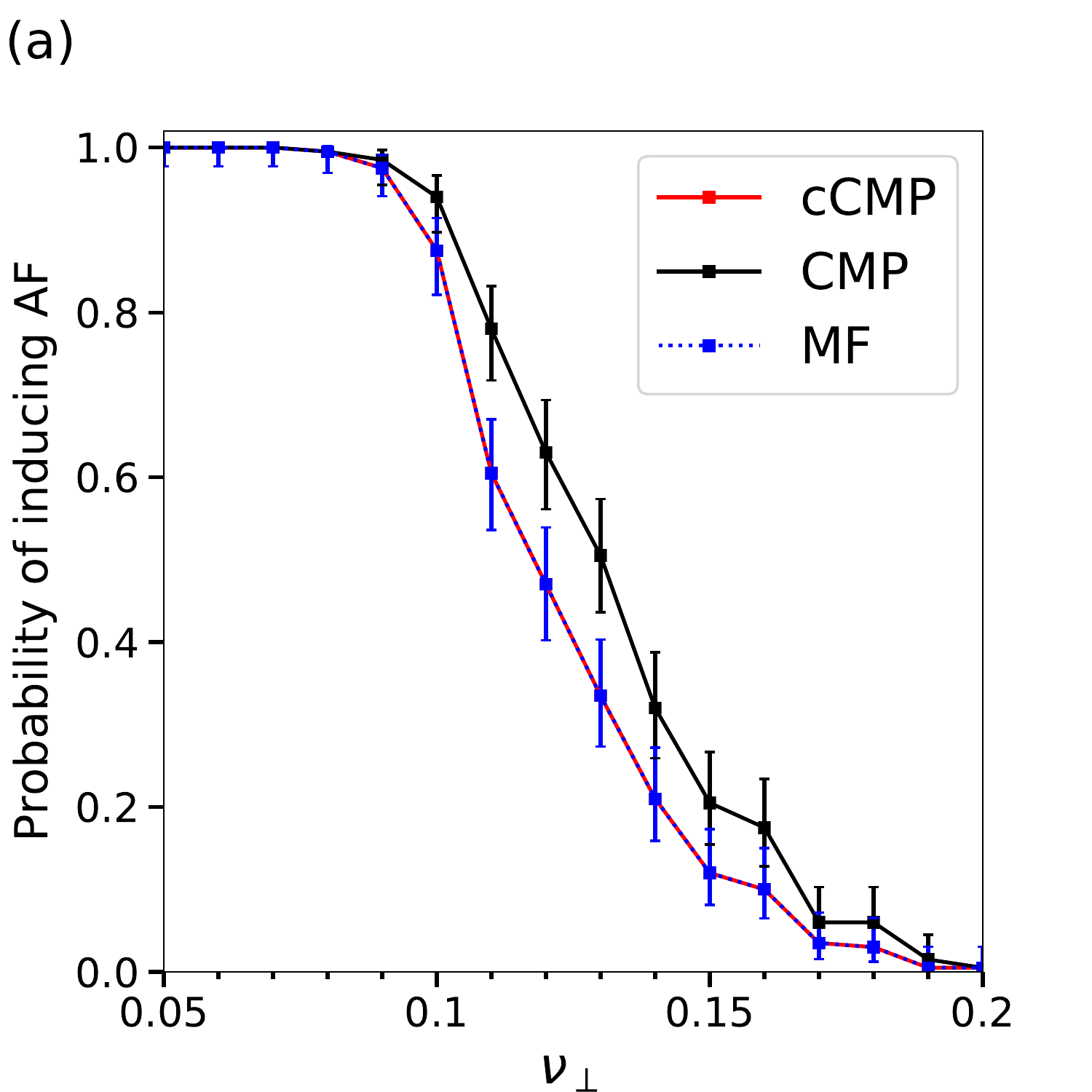}
\includegraphics[scale=0.4]{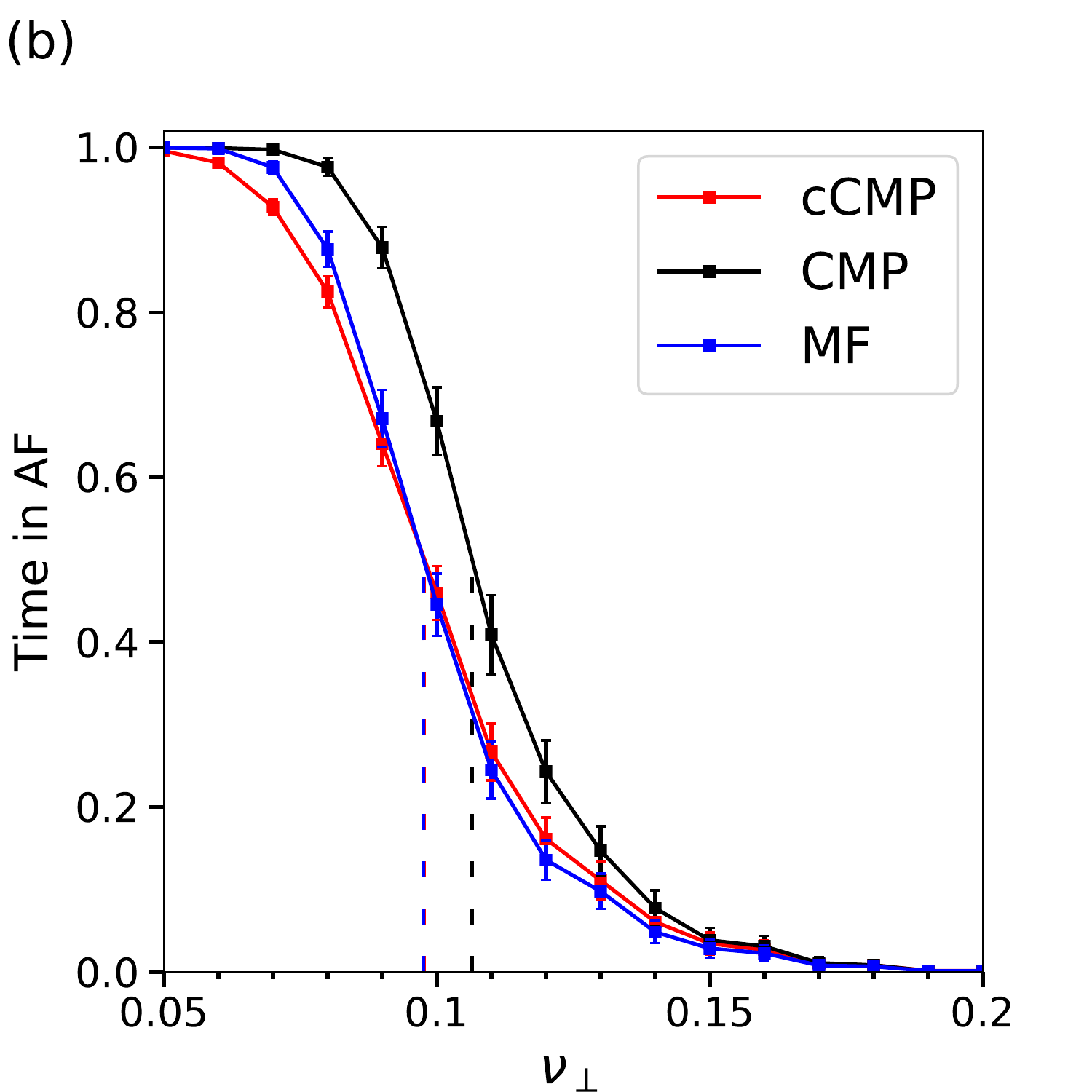}
\caption{Phase diagrams of (a) the probability of inducing AF and (b) the time in AF as a function of the fraction of transversal connections, $\nu_{\perp}$, for the CMP (black), cCMP (red) and MF (blue) models. Both statistics are significantly higher in the CMP model than in the cCMP and MF models. The probability of inducing AF is the same in both the cCMP and MF models since the number of particles in the MF model is taken from the cCMP model. Hence, if the cCMP has at least one critical structure, the MF model also has at least one particle. For both the cCMP and MF models, the probability of a structure/particle activating approaches $1$ as the simulation time is extended.
The time in AF in the MF model is slightly higher than in the cCMP model for $\nu_{\perp} \lessapprox 0.1$. In contrast, the cCMP model marginally overtakes the MF model for $\nu_{\perp} \gtrapprox 0.1$, see appendix~\ref{sec:MF_vs_cCMP}. Sharp transitions in the time in AF occur around the threshold values $\nu^{*}_{\perp} \approx 0.11$ (CMP, black dashed line), $\nu^{*}_{\perp} \approx 0.10$ (cCMP, red dashed line) and $\nu^{*}_{\perp} \approx 0.09$ (MF, blue dashed line).
 \label{fullphasediagram}}
\end{figure}

The probability of inducing AF and the time in AF are significantly higher in the CMP model than in the cCMP model, see Fig.~\ref{fullphasediagram}. This indicates that local regions with complex activation dynamics (e.g. multiple conduction blocks) provide a tangible contribution to AF emergence and maintenance. However, the cCMP and MF models do not show a significant difference in the time spent in AF, with the exception of small fluctuations above and below the critical coupling value, $\nu_{\perp}^{*}$.

The small differences between the MF and cCMP models can be understood as a consequence of the interaction between multiple active critical structures. By definition, the MF and cCMP models contain the same number of potential critical structures (or particles), $N$. In the MF model, particles activate independently of any other particles (with the exception of changes in the absolute rate of activation). Conversely, in the cCMP model high frequency waves emitted from a given re-entrant circuit may suppress the activation of new re-entrant circuits. Therefore, at low $\nu_{\perp} < \nu_{\perp}^{*}$ where there are many simple critical structures, the time spent in AF is slightly higher in the MF model than in the cCMP model since the presence of multiple simultaneous re-entrant circuits is not suppressed in the MF model, see appendix~\ref{sec:MF_vs_cCMP}.

In the reverse case at $\nu_{\perp} > \nu_{\perp}^{*}$, the cCMP slightly exceeds the MF due to slight differences in the rules of activation and deactivation. More precisely, a critical structure only has one opportunity to activate, or deactivate, in a given pacing cycle with probability $\epsilon$ in the cCMP model. However, particles can activate in the MF model every timestep with probability $\epsilon/T$. This results in the cCMP model spending marginally longer in AF if the number of potential simple re-entrant circuits is small, see appendix~\ref{sec:MF_vs_cCMP}.

\subsection{Critical Structures With Asymmetric Activation}

Since the spatial components of the CMP model do not account for the difference in the time spent in AF between the MF and CMP models, we note that the only remaining difference between the cCMP and CMP models is the distribution of conduction blocking nodes. By construction, the cCMP and CMP models contain the same number of simple critical structures. Therefore, some higher order critical structures must exist which rely on conduction blocking nodes which cannot form critical circuits by themselves, but which enhance the time spent in AF. 

Returning to Fig.~\ref{modeldynamics}, we draw particular attention to subfigures (b), (c), (d) and (f). In Fig.~\ref{modeldynamics}(b), we see an isolated number of very short events. Given the duration of the observed AF events, let us assume these dynamics are driven by a simple re-entrant circuit. With $T=220$, $\tau=50$ and $\epsilon=0.05$, the time for a single inactive re-entrant circuit to activate is approximately 4400 time steps. Similarly, once active, the time for a single active re-entrant circuit to deactivate is approximately 1000 time steps. This implies that, on average, from a single simple re-entrant circuit we expect to see a transition from sinus rhythm to fibrillation and back to sinus rhythm in approximately $5400$ time steps. For a simulation lasting $10^6$ time steps, this suggests a single simple re-entrant circuit might generate on the order of $\sim 185$ individual fibrillatory events. Such behaviour is visible in Fig.~\ref{modeldynamics}(c), but Fig.~\ref{modeldynamics}(b) only exhibits 5 events during this time span. This suggests that, although the probability of an active re-entrant circuit terminating is consistent with the presence of a simple re-entrant circuit, the probability of activating a re-entrant circuit is significantly suppressed. This implies the critical structure present in Fig.~\ref{modeldynamics}(b) requires multiple successive failures of conduction blocking nodes to activate, but only one to fail once active for the circuit to terminate. Examples of complex structures with these properties are described in Fig.~\ref{structuretypes}.

We now consider Fig.~\ref{modeldynamics}(d) where we observe a small number of isolated fibrillatory events similar to those shown in Fig.~\ref{modeldynamics}(b), implying the presence of a limited number of re-entrant circuits. However, unlike Fig.~\ref{modeldynamics}(b), the lifetimes of individual fibrillatory events in Fig.~\ref{modeldynamics}(d) are signficantly longer than the $\sim 1000$ time steps predicted for a single simple re-entrant circuit. This suggests that the re-entrant circuit in Fig.~\ref{modeldynamics}(d) has both a suppressed activation rate, and a suppressed deactivation rate relative to a simple re-entrant circuit. In these cases, multiple re-entrant circuits must fail simultaneously for fibrillation to be terminated, extending the lifetime of individual episodes. If we assume that the re-entrant circuit in Fig.~\ref{modeldynamics}(d) requires two successive failures for the circuit to terminate, this would imply an average fibrillatory event duration of $\sim 20000$ time steps. This demonstrates that a range of different re-entrant circuits can exist, at the same set of model parameters, which result in a spectrum of AF event durations. 

The example given in Fig.~\ref{modeldynamics}(d) requires two successive (in the same activation cycle) failures of conduction blocking nodes. However, we observe that in some cases, such as Fig.~\ref{modeldynamics}(h) activity persists for durations approaching $10^6$ time steps. For fibrillatory events to last this duration, circuits must form which require more than two successive failures of conduction blocking nodes to terminate.

\begin{figure}
\centering
\includegraphics[width=8.5cm, height = 8 cm]{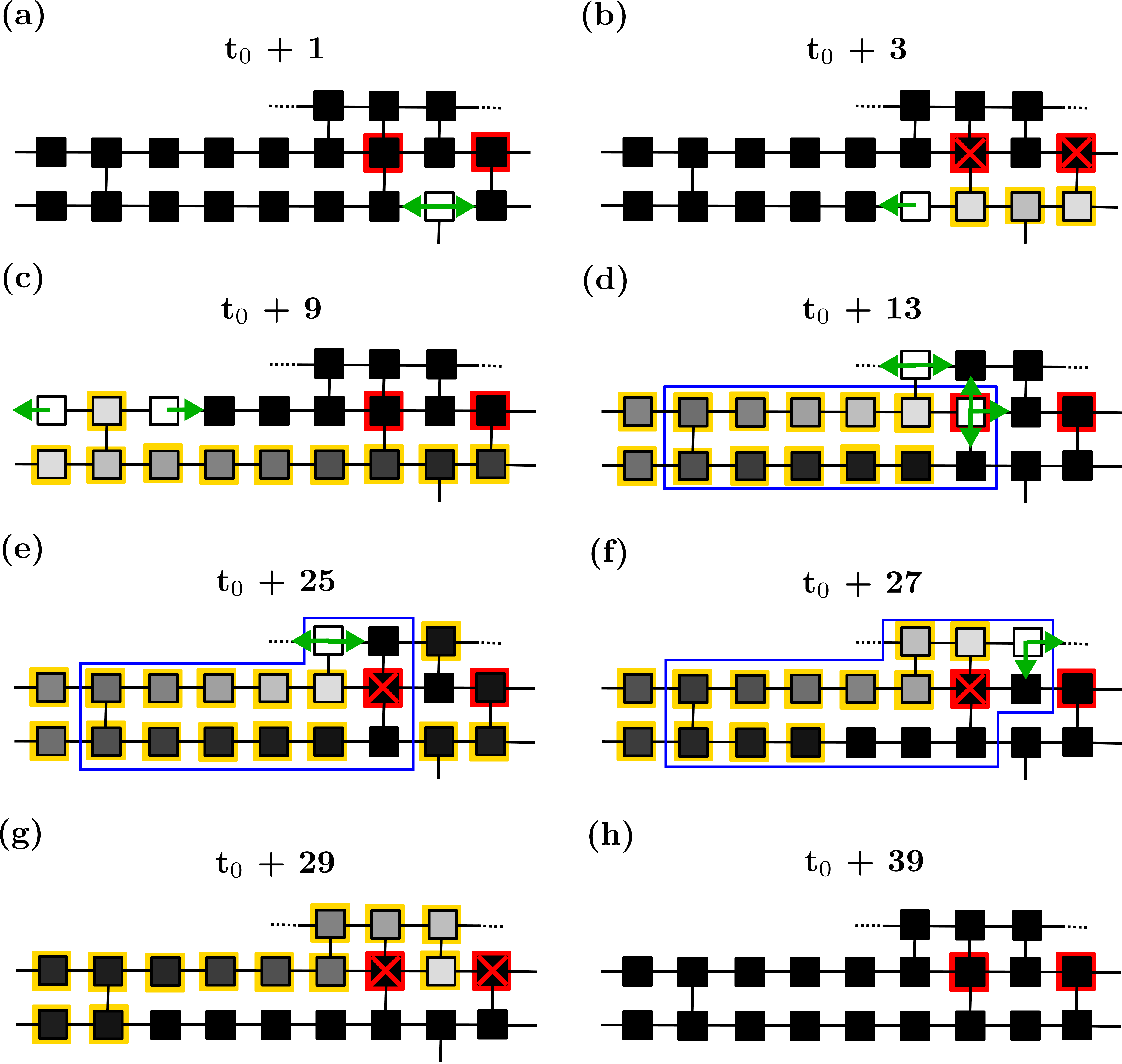}
\caption{
The formation of a complex re-entrant circuit with asymmetric activation and deactivation rates. 
(a) An excitation (white cell; not from sinus-rhythm) emerges from below the part of the CMP lattice in question, and propagates in the direction indicated by the green arrows. (b) The excitation is terminated by two conduction blocking nodes (red crosses), but the excitation successfully continues to the left of the lattice region. Refractory cells are shown in grey-scale with a yellow border. (c)-(d) The excitation branches to the adjacent fibre, propagates to the right, and branches back to the lower fibre forming a re-entrant circuit (blue box). (e)-(h) The re-entrant circuit is terminated from the successive failure of four conduction blocking nodes. Hence, this complex circuit requires two failures to initiate and four failures to terminate, resulting in long, persistent AF episodes. The full evolution of this structure is shown in the supplementary material with an accompanying video \cite{SM}.}
\label{fig:complex1}
\end{figure}

\begin{figure}
\centering
\includegraphics[width=8.5cm, height = 13 cm]{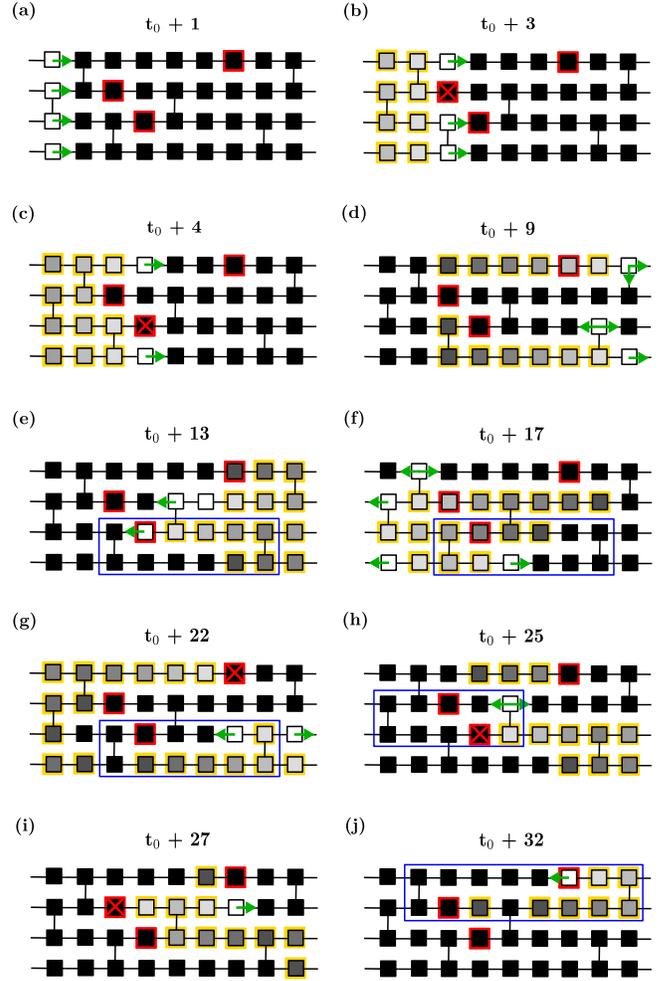}
\caption{The coupling of multiple re-entrant circuits such that the termination of one circuit immediately activates a secondary circuit. This coupled behaviour suppresses the probability that activity terminates in the CMP model. Colours as described previously. (a) A wavefront enters a region of the CMP model from sinus rhythm. (b)-(f) Two successive failures of conduction blocking nodes result in the initiation of a re-entrant circuit in the lower two fibres. (g) Activity propagating through the top fibre is blocked by the failure of a conduction blocking node. (h)-(i) The failure of both conduction blocking nodes in the central fibres allows the wavefront to re-enter the top fibre, forming a new re-entrant circuit. In this process, the termination of the initial re-entrant circuit has initiated a secondary re-entrant circuit, rather than restoring sinus rhythm. In total, two cell failures initiated the re-entrant activity, but activity has not been terminated after three additional cell failures. The full evolution of this structure is shown in the supplementary material with an accompanying video \cite{SM}.}
\label{fig:complex2}
\end{figure}

Inspecting the CMP model, we identify two structural mechanisms by which re-entrant circuits can form with asymmetric activation rates such that the probability of entering AF exceeds the probability of returning to sinus rhythm once AF has been initiated. The two mechanisms are as follows: (1) Self-contained critical structures where the number of cell failures required to initiate the structure is less than the number of failures required to terminate the structure. An example of such a structure is shown in Fig.~\ref{fig:complex1}. (2) Coupled critical structures whereby a fibre can be shared between multiple possible adjacent sub-structures such that when a re-entrant circuit has formed, its termination immediately initiates a new re-entrant circuit in a neighbouring fibre. Depending on the structures coupled, these circuits can require a vast number of cell failures for the activity to be terminated and for sinus rhythm to be restored. An example of such a structure is shown in Fig.~\ref{fig:complex2}.

In Fig.~\ref{fig:complex1}, a critical structure is initiated (not from sinus rhythm), which requires the successive failure of two conduction blocking nodes to activate. However, once active, four successive failures are requires to terminate the re-entrant circuit. This is because even if a single conduction blocking node fails to excite, secondary pathways exist such that re-entry can move around the conduction blocking region unimpeded. Only if all secondary pathways are blocked does the circuit terminate.

In Fig.~\ref{fig:complex2}, individual critical structures are coupled by sharing common segments. While an individual fibre may not have a higher probability of activation than termination, the coupled structures have the property that if one structure is terminated from a succession of conduction blocks, one of the coupled critical structures in the adjacent fibres can immediately initiate. Note, the probability of any of the sub-structures terminating is not fixed. Some critical structures may require only one cell failure to terminate, whereas other structures may persist in the presence of multiple successive cell failures. Through the inspection of many instances of the CMP model, there does not appear to be any clear limit to how many individual subregions can be coupled together in such a way that the termination of a given sub-structure doesn't initiate a new re-entrant circuit. If a large enough number of critical structures are coupled, and if some of those critical structures have asymmetric activation rates as shown in Fig.~\ref{fig:complex1}, individual fibrillatory events can become, to all practical purposes, indefinite. In a sample of one hundred simulations at very low coupling, $\nu_{\perp} = 0.05$, every simulation is observed to enter AF, and of those simulations, not a single case returns to sinus rhythm within $10^{6}$ time steps. Such observations suggest that, particularly at low coupling, coupled structures with very low termination probabilities dominate the CMP model.

\begin{figure}
\centering
\includegraphics[width=7.0cm, height = 7cm]{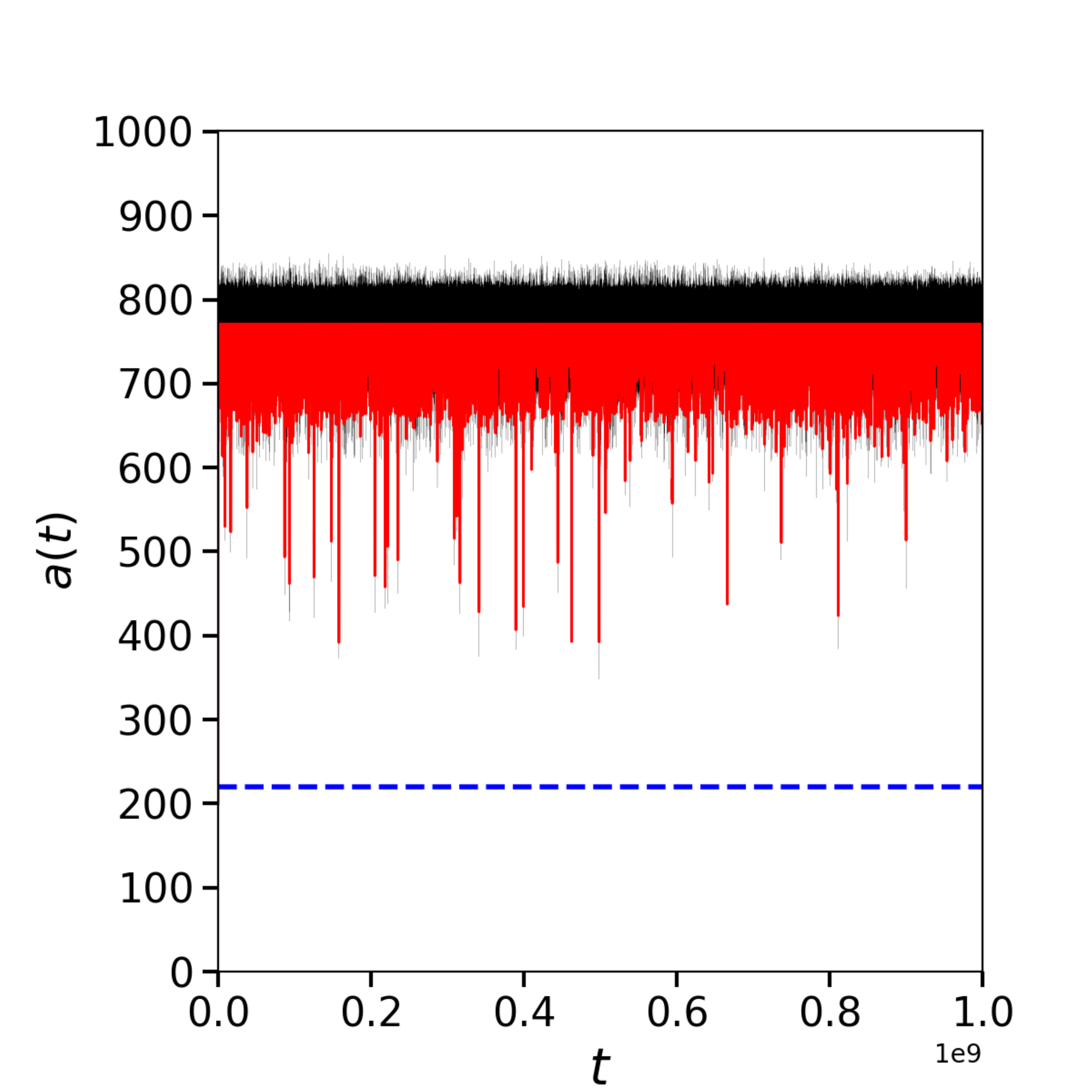}
\caption{A simulation of the CMP model for $10^9$ timesteps at $\nu_{\perp} = 0.05$. Persistent AF is maintained throughout the simulation. Converted into real time, this corresponds to an AF event of at least 1 month in duration.
\label{longexp}}
\end{figure}

To emphasise the extent to which these mechanisms can enhance the probability of persistent AF, Fig.~\ref{longexp} shows the number of active cells for a simulation which requires a small number of conduction blocks to activate, but at least 5 successive failures to terminate. In such situations, fibrillation can be maintained for durations in excess of $10^9$ time steps. This indicates that with simple dynamics and rules, the CMP model is capable of exhibiting fibrillatory events which, if converted into real time, can span anywhere from seconds to months for the same set of model parameters.

\begin{figure}
\centering
\includegraphics[width=9.0cm, height = 9.0cm]{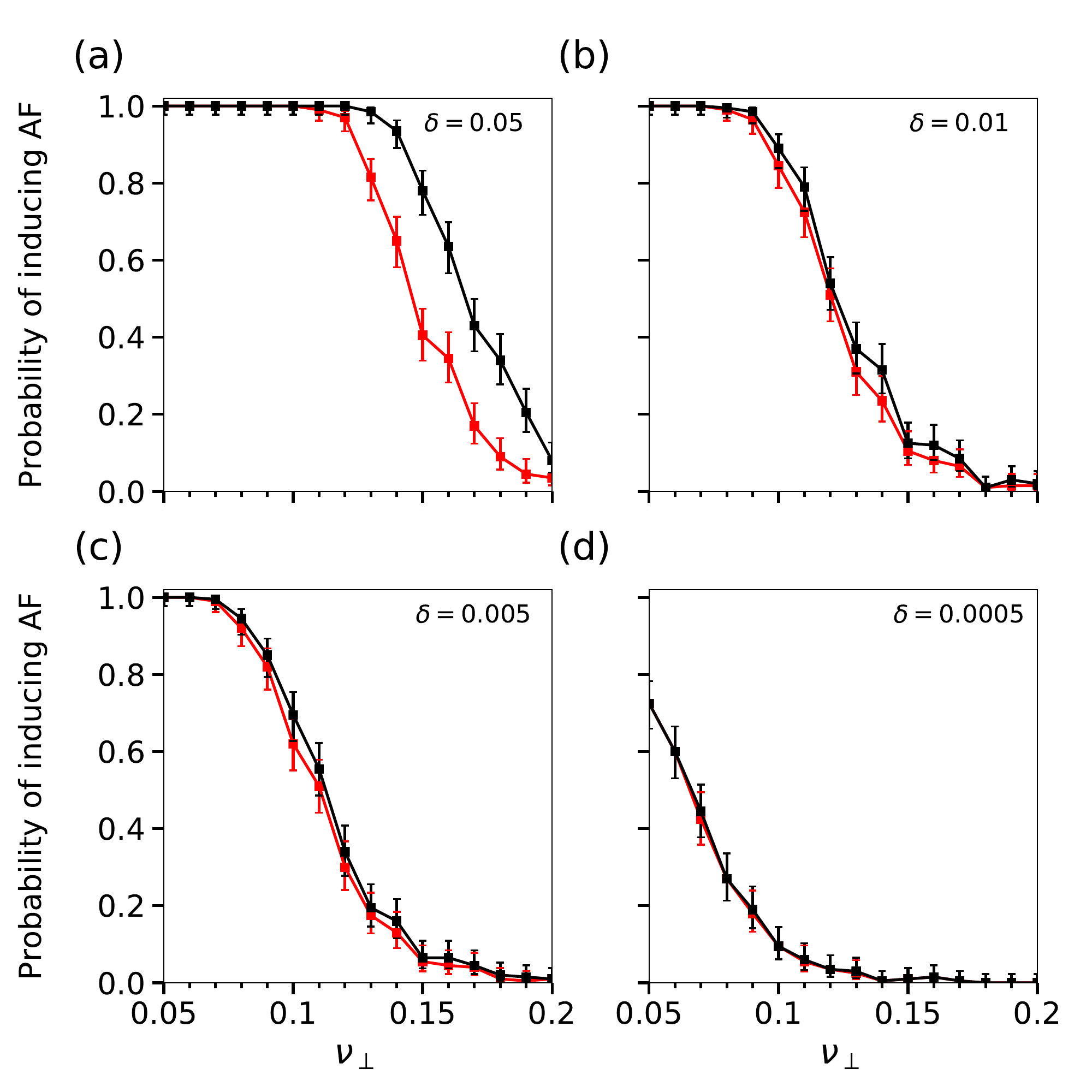}
\caption{Phase diagrams for the probability of inducing AF in the CMP model (black) and the cCMP model (red) as a function of the coupling probability, $\nu_\perp$. Each subfigure is for a different value of $\delta$, the fraction of nodes that are susceptible to conduction block. In general, the figures indicate that reducing $\delta$ reduces the risk of AF. This is consistent with the theoretical risk shown in Eq.~\eqref{theorycurve}. Additionally, the figures indicate that as $\delta$ is reduced, the excess risk of entering AF in the CMP model relative to the cCMP model reduces until both the CMP and cCMP models collapse onto the same curve. This indicates that at high $\delta$, the density of conduction blocking nodes in the lattice is sufficiently high such that complex re-entrant circuits, see Fig.~\ref{fig:complex1} and Fig.~\ref{fig:complex2}, have a noticeable contribution to the risk of entering AF in the CMP model. However, as the density of conduction blocking nodes is reduced, the probability that two conduction blocking nodes are sufficiently close to form a complex re-entrant circuit becomes vanishingly small. As a result, the risk of entering AF is dominated by simple re-entrant circuits that require only a single cell failure to induce AF. Consequently, the risk of entering AF in the model coincides with the cCMP risk where all critical structures are simple re-entrant circuits.}
\label{ChangeDelta}
\end{figure}

A common feature in the complex critical structures we have identified is that their activation and deactivation relies on the failure of multiple conduction blocking nodes within a small critical region. This suggests that the formation of these structures should be dependent on the local density of conduction blocking nodes, $\delta$. In Fig.~\ref{ChangeDelta} we plot the phase diagrams for the probability of inducing AF in the cCMP and CMP models at different values of $\delta$. As $\delta$ increases (decreases) the spatial density of conduction blocking nodes in the cCMP and CMP models increases (decreases). Figure~\ref{ChangeDelta} demonstrates that at large $\delta$, the CMP model is significantly more likely to enter AF than the cCMP model. However, this difference disappears as the spatial density of conduction blocking nodes is lowered. This indicates that the dominant contribution to the difference in the probabilities of inducing AF comes from these special re-entrant circuits with asymmetric activation and termination rates as opposed to the special case shown in Fig.~\ref{criticalstructure} where only one failure is required to initiate and terminate a re-entrant circuit.

\section{\label{CMP_Context} The CMP Model in Context}

The CMP model is a discrete, cellular automaton model of AF where the myocardium is treated as a discrete structure. Discrete models of AF have a long history in cardiac electrophysiology modelling, including the first quantitative computational study on AF by \citeauthor{moe1964computer} \cite{moe1964computer}. Despite their popularity in the early days of computational cardiac electrophysiology, in recent years, reaction-diffusion models, where the myocardium is treated as a continuum, have superseded discrete models in popularity \cite{clayton2011,Niederer2019}. However, discrete models have remained popular specifically for studying how the accumulation of fibrosis in the atria effects the initiation and maintenance of AF. In our opinion, such studies are particularly important because, although existing reaction-diffusion models on a continuous myocardium are highly effective at simulating functional reentry and testing the effect of prospective pharmacological treatments, without discontinuities in the microstructure of the myocardium, these models cannot effectively study the emergence of micro-anatomical reentry where reentry is not functional, nor anchored to macroscopic lesions in the tissue. 

Dynamics in discrete models of AF are either modelled as a discrete rule based process, corresponding to a cellular automaton, or using the reaction-diffusion equations typically used in continuous models of cardiac electrophysiology. In the original Moe model \cite{moe1964computer}, dynamics were simulated using a cellular automaton on a hexagonal lattice. \citeauthor{moe1964computer} focus on studying the role of spatial inhomogeneity in the action potential duration (the refractory period), and how the degree of inhomogeneity effects the emergence of turbulent propagation in AF, supporting the multiple wavelet hypothesis previously proposed by Moe in \cite{moe1959}. However, the paper predominantly focuses on the maintenance of AF with initiation being induced from a burst pacing protocol. Additonally, although the myocardium is treated as discrete, neighbouring nodes in the model are not explicitly decoupled (simulating the action of fibrosis), as is done in the CMP model. As a result, although the work by \citeauthor{moe1964computer} was pivotal in giving a theoretical explanation for AF maintenance from the multiple wavelet hypothesis, the lack of discontinuities in the microstucture of the simulated myocardium mean that the model does not consider the initiation of AF from micro-anatomical re-entry as is done in the CMP model. 

The more recent discrete models of AF have, on the whole, included discontinuities in the microstructure of the simulated myocardium \cite{alonso2013,alonso2016,lin2017,kazbanov2016}. Of particular interest to the CMP model is the work by \citeauthor{alonso2013} \cite{alonso2013} where the authors study the emergence of AF from the accumulation of fibrosis in a local region of the myocardium and across the myocardium as a whole. In particular, the authors associate the risk of inducing AF with the approach from above of the site (or bond) occupation probability towards the percolation threshold, and more specifically, how the probability of re-entry is associated with the emergence of insulating clusters of fibrosis with a linear dimension greater than a critical value $\phi$. This result is in line with our recent work on the three-dimensional extension of the CMP model where we discuss how the formation of insulating clusters effects the dynamics of transmural reentry  and the emergence of simulataneous focal breakthrough drivers and re-entrant drivers \cite{falkenberg2019}. These results are important for the understanding of micro-anatomical reentry in AF because it explicitly associates local decoupling in the myocardium with the emergence of local drivers in AF, and it gives plausible explanations for why, when the density of fibrosis is too high, local ablation may struggle to successfully terminate AF. \citeauthor{alonso2016} have also extended their model to three dimensions to study how the thickness of tissue effects the probability of AF initiation as fibrosis accumulates \cite{alonso2016}. Additionally, the authors investigate how re-entrant dynamics in fibrotic regions may explain the complex fractionated atrial electrograms observed clinically near infarct regions. Similar results have been observed in other simple discrete models studying the reentry dynamics that emerge from specific fibrosis patterns \cite{vigmond2016}.

The primary difference between the work by \citeauthor{alonso2013} and our work on the CMP model is the methodology for simulating dynamics on the myocardium. The CMP is a cellular automaton with discrete rule based propagation of action potential. Contrastingly, \citeauthor{alonso2013} keep the reaction-diffusion model framework common in other models of cardiac electrophysiology, simulating the propagation of action potentials using the Fenton-Karma model \cite{fenton1998}. The Fenton-Karma model is a relatively simple phenomenological model for simulating wavefront propagation on cardiac tissue focused on qualitatively reproducing the collective behaviour of key ion channels. This neglects much of the detail present in more complex reaction-diffusion models \cite{clayton2011}, although unlike the CMP model, the Fenton-Karma framework does account for action potential and conduction velocity restitution. While the importance of action potential and conduction velocity restitution is well established for functional re-entry, it is not absolutely clear how important these features are to the emergence of micro-anatomical re-entry. In general, both features are known to be proarrhythmic, so the exclusion of these features in the CMP model gives credence to the idea that AF can emerge due to the accumulation of fibrosis only, without the need for additional proarrhythmic features. However, we acknowledge that despite this, the omission of these features does limit the CMP model to qualitative, rather than quantitative predictions, although this is a limitation also acknowledged by \cite{alonso2013} when using simple reaction diffusion models like the Fenton-Karma model. 

One of the benefits of omitting rate dependent effects in the CMP model is that the simplicity of this framework allows for an in depth analytical treatment of the risk of AF initiation as a function of the density of fibrosis in a local region. A result of such an analytical treatment is that, although we do not include action potential or conduction velocity restitution, the theoretical form of risk of inducing AF does allow us to predict how the CMP model would behave if these features were included. We can rewrite the risk of inducing AF as a function of a parameter $\gamma$,
\begin{equation}\label{theorycurve_expanded}
    R = 1 - \left[1 - ( 1 - \nu_{\perp})^\tau \right]^{\delta L^2} = 1 - \gamma^{\delta L^2},
\end{equation}
where $\gamma = 1 - ( 1 - \nu_{\perp})^\tau$, corresponding to the probability that a re-entrant circuit does not exist at specific lattice site in the CMP model. The exponent $\delta L^2$, the average number of conduction blocking nodes, indicates that the risk of AF is dependent on the tissue dimensions and the susceptibility of nodes to conduction block. From Eq.~\eqref{theorycurve_expanded}, it is clear that maximising the risk of AF is achieved by minimising $\gamma$, which in turn is achieved by minimising $\nu_{\perp}$ and/or $\tau$. Noting that $\nu_{\perp}$ is a probability bounded by $0 \leq \nu_{\perp} \leq 1$, and excluding the cases of a fully coupled or uncoupled lattice, an increase (decrease) in the coupling value, $\Delta \nu_{\perp}$, can always be cancelled out by a decrease (increase) in the refractory period, $\Delta \tau$, and vice versa, such that there is no change in the risk of inducing AF. That means that for any refractory period, given a sufficiently small (non-zero) value for $\nu_{\perp}$, it is always possible to induce AF in the CMP model. 

Recall that in the CMP model, a re-entrant circuit is induced from sinus rhythm in a local region from a single sinus wavefront. However, as noted previously, a change in the coupling value $\Delta \nu_{\perp}$, and likewise the local coupling value, can be counteracted by a suitable change in the refractory period $\tau$. That means that reducing the coupling value in a local region is equivalent to fixing the coupling value and reducing the refractory period, corresponding to action potential restitution. Likewise, rather than explicitly simulating action potential restitution, the equivalent effect is observed by reducing the coupling value $\nu_{\perp}$. 

Although the CMP model in its current form does not explicitly introduce spatial heterogeneity in $\nu_{\perp}$, individual bonds in the model are filled probabilistically meaning that in a local region, the actual coupling value $\tilde{\nu}_{\perp}$ will be fluctuate around $\nu_{\perp}$ with standard deviation $\sigma(\nu_{\perp})$. Consequently, regions with a local coupling value below the average are more likely to harbour a re-entrant circuit than other local regions, and these regions will dominate the fibrillatory dynamics in the CMP model as previously shown in \cite{kishanthesis,manani2016}. Explicitly introducing spatial heterogeneity in the coupling values would not change these results, but rather, would only change the value of $\sigma(\nu_{\perp})$ and the resulting local coupling value $\tilde{\nu}_{\perp}$. Note, that this implies that if the spatial heterogeneity is increased, the probability of regions with particularly low coupling relative to the mean increases, and hence, since these low coupling regions dominate the risk of fibrillation, higher spatial heterogeneity in the accumulation of fibrosis will result in a greater risk of inducing fibrillation. This argument is consistent with the results of another discrete model of AF by \citeauthor{kazbanov2016} \cite{kazbanov2016} where the authors study the initiation of AF in fibrotic conditions following similar methods to those applied in \cite{alonso2013}. The authors demonstrate that AF risk increases with spatial heterogeneity and they also note that the regions which dominate the risk of AF induction are those with the minimal local coupling value. 

A similar argument regarding the role of spatial heterogeneity in the fibrosis distribution can be applied to conduction velocity restitution. In the model, $\gamma$ is written as a function of the refratory period $\tau$ rather than the refractory wavelength $\lambda(\tau,v) = \tau v$, where $v$ is the conduction velocity, since in the cellular automata framework the conduction velocity is 1 node per timestep. However, if variable conduction velocity was included in the CMP model, we can rewrite the probability of a re-entrant circuit not existing as $\gamma = 1 - ( 1 - \nu_{\perp})^\lambda = 1 - ( 1 - \nu_{\perp})^{\tau v}$. Written in this form, it is clear that reducing the conduction velocity has an equivalent effect on the risk of forming a re-entrant circuit as reducing the refractory period.  

In summary, although the CMP model does not include several important details of cardiac electrophysiology, such as action potential restitution, conduction velocity restitution, and spatial heterogeneity in fibrosis, the simplicity of the CMP framework allows an analytical treatment that suggests how these more complex features would effect the probability of forming re-entrant circuits. We find that both action potential and conduction velocity restitution increase the probability of forming a re-entrant circuit. 

Apart from aiding in the derivation of analytical results, simulating dynamics in the CMP model using a rule based cellular automata framework has significant computational efficiency benefits that enable the study of rare events in cardiac electrophysiology that may only take place clinically on the timescale of hours to days as highlighted in another recent cellular automata model \cite{lin2017}. This includes studying the transitions between paroxysmal and persistent AF which we focus on here. While there is interest in studying these dynamics in detailed, biophysical reaction diffusion models using realistic topologies, these models cannot feasibly study fibrillation dynamics over long enough time scales to investigate these questions. More precisely, the simulations described in this manuscript last for up to $10^9$ time steps, corresponding to approximately a month in real time. Such time scales are out of reach for current biophysically detailed reaction-diffusion models, struggling to exceed 60 seconds in real time \cite{lin2017}.

Despite these benefits, it is critically important to stress that the efficiencies of the CMP model come with significant limitations and as such, the results of the CMP model are principally valuable for generating new hypotheses to be tested in more complex models of fibrillation, or clinically. The results presented should be understood qualitatively; the event durations presented do not necessarily reflect the events that would be observed in a clinical case of AF.

The key limitations of the CMP model include the following: (1) The use of a square 2d lattice restricting the degrees of freedom in which excitations can move. In the case of fluid flow, it is known that a square lattice is not able to conserve vorticity in lattice-boltzmann simulations and that a hexagonal lattice is preferable \cite{wolf2005}. This is partially responsible for the inability of the CMP model to maintain fibrillation with spiral-wave re-entry. Adapting the CMP model to a hexagonal lattice to enable spiral-wave re-entry is currently underway. (2) A number of electrotonic effects are excluded including APD and CV restitution, and the formation of alternans, beat to beat fluctuations in the length of the action potential. It is not clear whether alternans may effect the initiation and maintenance of re-entrant circuits. (3) The fibres in the CMP model lie along the same axis, with cells perfectly coupled within a given fibre. In real atrial tissue, fibres meander with complex orientations. This may effect how neighbouring re-entrant circuits interact. Note, the use of a real atrial fibre structure has been applied to the CMP model in \cite{falkenberg2019}. (4) The extended simulations in the CMP model do not consider the role of electrical or structural remodelling. Given that some of the simulations in the CMP model represent approximately 1 month in real time, clinically we may expect to observe extensive remodelling in the myocardium, although that is neglected in the CMP model. The role of structural remodelling is, to an extent, considered in \cite{manani2016}. Despite these limitations, investigations into the persistence of AF arising from the initiation and termination of micro-anatomical re-entrant circuits are, to the best of our knowledge, wholly novel to the CMP model.

\section{\label{DisConOut} Discussion, conclusion \& outlook}

The persistence of AF is one of the key factors determining the likelihood of a successful ablation \cite{calkins2017}. However, at a microstructural level, it is not clear what determines whether a patient will exhibit paroxysmal or persistent AF \cite{sugihara2015,kottkamp2013}. Better understanding the causes of persistent AF may, in the future, improve the success rate of ablation or inspire new potential treatment methods.

The CMP model is a simple, percolation based model of AF where re-entrant circuits form when adjacent muscle fibres decouple \cite{christensen2015}. The model is not a fully realistic representation of atrial electrophysiology. However, the model, and its extensions, have offered explanations for a number of key observations from clinical AF research. This includes the diversity of AF persistence at comparable levels of fibrosis \cite{manani2016}, the distribution of re-entrant circuits in the left and right atria \cite{falkenberg2019}, and the observation that re-entrant circuits preferentially form near the endocardium (inner heart wall) rather than the epicardium (outer heart wall) in paroxysmal AF \cite{falkenberg2018,hansen2015}. Additionally, the 3D extension of the CMP model suggests a natural explanation for the lowering success rate of ablation as AF becomes more persistent \cite{falkenberg2019}. Despite these findings, the precise dynamics at the microscopic level of the CMP model were not fully elucidated -- until now it was not clear how the model is capable of showing the full diversity of AF persistence.

In this paper, we have focused on better understanding the microscopic dynamics of the CMP model, specifically with the aim to understanding which microscopic interactions are responsible for the emergence of persistent AF. By disecting the model into its constituent parts, we have shown how the formation of complex re-entrant circuits which have a large probability of activating, but a significantly smaller probability of terminating accounts for the difference between the CMP model and the MF/cCMP models. Once activated, these drivers exhibit a wide spectrum of AF event lifetimes from, in real time, a few seconds to months. This spectrum reflects the broad range of AF subtypes exhibited by the model at the same model parameter values, from paroxysmal to persistent AF. 

To identify the emergence of persistent re-entrant circuits, we first derived a mean-field model of AF, neglecting the spatial components of the model and interactions between coexisting drivers. Mean-field approaches are well established in physics for simplifying the study of high dimensional random processes \cite{ChristensenMoloney:2005}. These models have been used extensively across numerous interdisciplinary field including in the study of epilepsy in neuroscience \cite{soltesz2011,rodrigues2009,marten2009}. Given the qualitative similarities between epilepsy and cardiac fibrillation, it is surprising that mean-field models are not widely used in computational cardiac electrophysiology.

Our mean-field model demonstrates that the essential features of AF remain if spatial structure and driver interactions are neglected. However, the mean-field model significantly underestimates the time spent in fibrillation relative to the CMP model, and it does not explain the emergence of persistent AF. Only by reintroducing spatial structure can these observations be explained. 

Re-introducing spatial structure in the controlled CMP model where we carefully control the initiation of simple re-entrant circuits, we have shown that the density of conduction blocking nodes plays a key role in the time the CMP model spends in AF. At high densities, the CMP model spends significantly more time in AF than the cCMP model. However, as the density of conduction blocking nodes is reduced only simple structures, like those found in the cCMP model, can form in the CMP model. As a result, the time in AF converges for the cCMP and CMP models.

%\subsection{Limitations}

It is important to stress that the results presented here are for a highly simplified physics model of AF. The scope of the CMP model is highly specific, focusing on the emergence of re-entrant circuits from the accumulation of fibrosis in the atria. We use cellular automata in our modelling approach which limits the realism of the dynamics in our model, but which recent research has suggested may be preferable to detailed continuous models when studying the effects of local heterogenity in the cardiac microstructure \cite{gokhale2017}, e.g. due to fibrosis. The model is both structurally and topologically simplified - we do not account for variation in fibre orientation (as in \cite{falkenberg2018}), nor do we consider the real topology of the atria (as in \cite{falkenberg2019}). Additionally, we do not consider variations in the action potential which are present in models which study the ionic currents across gap junctions \cite{clayton2011}. 

However, the value of such a simple model should not be underestimated. Cellular automata are very computationally efficient, allowing for a statistical analysis not easily achieved in more complex models. Likewise, the model has very few key parameters, with re-entrant circuits emerging, and the diversity of AF persistence being explained, by the variation in a single coupling parameter, $\nu_{\perp}$. This gives clarity to any results, avoiding ambiguity as to which model features are responsible for the emergence and maintenance of AF. Finally, although the CMP model itself may not represent a fully realistic atrial electrophysiology, the extensions of the CMP model to 3D and to a real topology are bringing the model closer to clinical relevance. However, naturally, these adaptations complicate model analysis. Hence, understanding the dynamics of simple models is essential to fully understanding the behaviour of the more complicated adaptations for which the CMP model is a precursor. Models such as the CMP model have significant potential in hypothesis generation and will play an increasingly important role in bridging the gap between clinical and computational electrophysiology -- such work is already going on in the ElectroCardioMaths centre at Imperial College London, as well as in other groups.

Given that ``AF begets AF'', finding ways to treat and prevent persistent AF is a key priority in AF research. In this paper, we have studied the microstructural basis for the emergence of persistent AF in the Christensen-Manani-Peters model. We have shown that persistent AF can arise from the formation of re-entrant circuits with an asymmetry in their probability of activation relative to the probability of termination. These circuits, once active, may drive AF for anywhere from a few seconds to months. 
%We have shown that the probability of forming these secondary re-entrant circuits (and equivalently the probability of persistent AF) increases as the total level of coupling in the model decreases. This is in line with clinical modelling suggesting that the formation of re-entrant drivers in fibrotic border zones perpetuate persistent AF \cite{zahid2016}. 

Future work should focus on validating the results obtained here in structurally realistic models of AF, derived from experimentally acquired fibre maps. If successful, this approach may suggest the regions of the atria most susceptible to the formation of persistent re-entrant circuits, and hence, may suggest suitable targets for ablation in persistent AF.

\section{Acknowledgements}

M.F. and A.C. gratefully acknowledge PhD studentships from the Engineering and Physical Sciences Research Council through Grants No. EP/N509486/1 and No. EP/L015129/1 respectively. N.S.P. acknowledges funding from the British Heart Foundation (Grant No. RG/16/3/32175 and Centre for Research Excellence RE/18/4/34215) and the National Institute for Health Research Biomedical Research Centre. K.C. and N.S.P. acknowledge funding from the Rosetrees Trust Grant A1173/M577.
\newpage

\appendix
\section{\label{CMF} Mean-Field Model of AF in Continuous Time}
The MF model can be extended to the continuous time case (cMF), providing us with a framework in which the time in fibrillation can be computed analytically. Let $\tilde{p}(k,t)$ be the probability of observing $k$ active simple re-entrant circuits at time $t$. When the interval between two consecutive time steps $\Delta t$ is sufficiently small (i.e.,  $\Delta t \to 0$), we have at most one event (activation or de-activation) per interval. In these settings, the dynamics of $\tilde{p}(k,t)$ are described by the following master equation
\begin{subequations}\label{mastereq}
	\begin{align}
	\frac{d\tilde{p}(k,t)}{dt} & = p(N-k+1)\tilde{p}(k-1,t) - p(N-k)\tilde{p}(k,t) \notag \\ &\hspace*{-0.5cm}  + q(k+1)\tilde{p}(k+1,t) - qk\tilde{p}(k,t) \quad \text{for $k > 1$} 
	\label{mastereqa}
	\end{align}
	where the first two terms are associated with an activation process $k-1 \mapsto k$ and $k \mapsto k+1$ transitions, respectively,
	while the last two are associated with $k+1 \mapsto k$ and $k \mapsto k-1$ transitions, respectively.
	Because the activation rate is different when the system has no active particles, see Eq.~\eqref{rates_a},
	we need to take special care of the $k=1$ and $k = 0$ cases.
	If the term $p_0$ represents the activation rate when the system has no active particles, then
	\begin{alignat}{2}
	\frac{d\tilde{p}(1,t)}{dt} =&\;\;  p_{0}N\tilde{p}(0,t) - p(N-1)\tilde{p}(1,t) \notag \\ 
	&\;+ 2q\tilde{p}(2,t) - q\tilde{p}(1,t), \hspace*{0.75cm} \text{for $k = 1$,}
	\label{mastereqb} \\
	\frac{d\tilde{p}(0,t)}{dt} = &  
	-p_{0}N\tilde{p}(0,t) + q\tilde{p}(1,t), \hspace*{0.5cm} \text{for $k = 0$.}
	\label{mastereqc}
	\end{alignat}
\end{subequations}
We enforce the boundary conditions that $\tilde{p}(k, t) = 0$ for $k < 0$ and $k > N$.
We will find the steady state solution $\tilde{p}(k) = \lim\limits_{t\rightarrow\infty}\tilde{p}(k,t)$
where the derivatives on the left-hand side of Eq.~\eqref{mastereq} are zero by the ansatz
\begin{equation}\label{ansatz}
\hspace*{-0.2cm}\tilde{p}(k) \!=\! \begin{cases}
A\binom{N}{k}\left(\frac{q}{p}\right)^{N-k} \!\!+ B\delta_{k,0} &\hspace*{-0.5cm}\quad \text{for $k \!=\! 0, 1,\ldots, N$}, \\
0 &\hspace*{-0.5cm}\quad \text{for $k \!<\! 0$ or $k \!>\! N$},
\end{cases}
\end{equation}
where $\delta_{i,j}$ is the Kronecker delta function. By inserting the ansatz into Eq.~\eqref{mastereqa}, we confirm that it solves the steady state equation for $k > 1$. However, in our case it simplifies further as $p = q = \epsilon/\langle \ell \rangle$, see Eq.~\eqref{rates_a}, so $q/p = 1$.
\begin{comment}
First, we check the proposed solution for \refeq{mastereqa}. For the first term we find that
\begin{equation}
\begin{aligned}
\begin{split}
& p(N-k+1)\tilde{p}(k-1,t) - p(N-k)\tilde{p}(k,t) \\
& = ApN!\left(\frac{q}{p}\right)^{N-k+1}\left(\frac{(N-k+1)}{(N-k+1)!(k - 1)!} - \left(\frac{p}{q}\right)\frac{(N-k)}{(N-k)!k!}\right)\\
& = ApN!\left(\frac{q}{p}\right)^{N-k+1}\left(\frac{1}{(N-k)!(k - 1)!} - \left(\frac{p}{q}\right)\frac{1}{(N-k-1)!k!}\right)
\end{split}
\end{aligned}
\end{equation}
while for the second term we observe that
\begin{equation}
\begin{aligned}
\begin{split}
& q(k+1)\tilde{p}(k+1,t) - qk\tilde{p}(k,t) \\
& = qN!\left(\frac{q}{p}\right)^{N-k}\left(\frac{(k+1)}{(N - k - 1)!(k+1)!}\left(\frac{p}{q}\right) - \frac{k}{(N-k)!k!}\right) \\
& = qN!\left(\frac{q}{p}\right)^{N-k}\left(\frac{1}{(N - k - 1)!k!}\left(\frac{p}{q}\right) - \frac{1}{(N-k)!(k-1)!}\right)
\end{split}
\end{aligned}
\end{equation}
so we require that
\begin{equation}
\begin{aligned}
\begin{split}
0 & = p\left(\frac{q}{p}\right)\left(\frac{1}{(N-k)!(k - 1)!} - \left(\frac{p}{q}\right)\frac{1}{(N-k-1)!k!}\right) \\ 
& + q\left(\frac{1}{(N - k - 1)!k!}\left(\frac{p}{q}\right) - \frac{1}{(N-k)!(k-1)!}\right) \\
& = \frac{q}{(N-k)!(k - 1)!} - \frac{p}{(N-k-1)!k!} + \frac{p}{(N - k - 1)!k!} \\
& - \frac{q}{(N-k)!(k-1)!}
\end{split}
\end{aligned}
\end{equation}
this confirms the ansatz \eqref{ansatz} for \eqref{mastereqa}.
\end{comment}

\begin{figure}
\centering

\includegraphics[width=7.0cm]{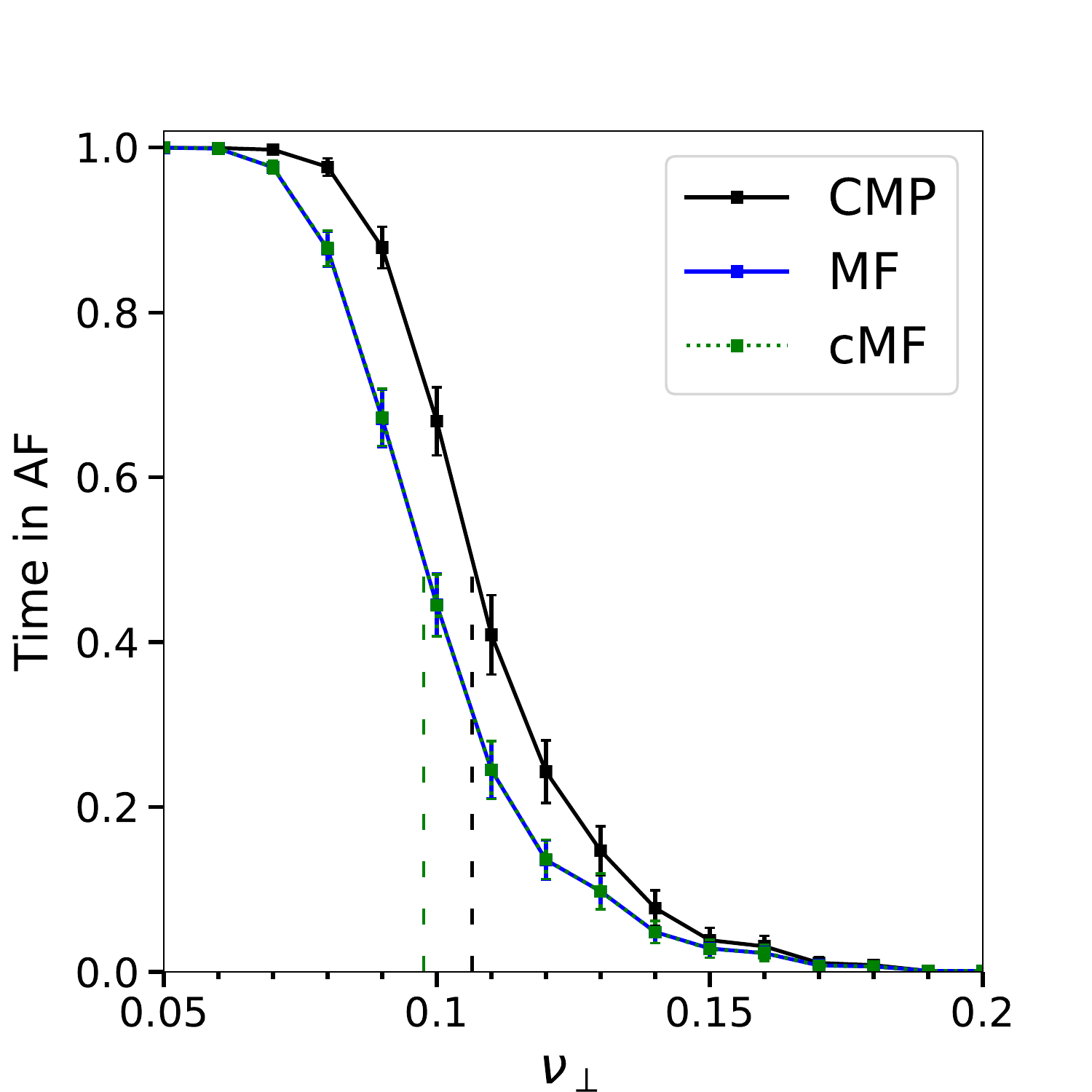}
\caption{Phase diagram of the time in AF as a function of the fraction of transversal connections $\nu_{\perp}$ for the CMP (black), the MF (blue) and the cMF (green) models. We use the parameters of the CMP model (i.e.,  $N, \epsilon$ and $\langle \ell \rangle$) to calibrate the cMF model and calculate the time in AF according to Eq.~\eqref{TAFC}. The phase diagrams for the MF and the cMF models are perfectly compatible. Both models significantly underestimate the time in AF with respect to the CMP model. Sharp transitions in the time in AF occur around the threshold values $\nu^{*}_{\perp} \approx 0.11$ (CMP, black dashed line), $\nu^{*}_{\perp} \approx 0.09$ (MF, blue dashed line) and $\nu^{*}_{\perp} \approx 0.09$ (cMF, green dashed line). \label{fig:cmf}}
\end{figure}

We can determine the two constants $A$ and $B$ by requiring that Eq.\eqref{ansatz} solves
Eqs.~\eqref{mastereqb}-\eqref{mastereqc} together with the normalization constraint:
inserting the ansatz into Eq.~\eqref{mastereqc}, recalling $p=q$, we find
\begin{align}
0 = -p_{0} A N -p_{0}NB +qAN
\label{ansatz3}
\end{align}
implying that
\begin{subequations}
	\label{Eq:AB}
	\begin{equation}
	B = A \left(\frac{p}{p_0} - 1\right).
	\label{Eq:AB1}
	\end{equation}
Note that $p = p_0 \Rightarrow B = 0$, that is, $\tilde{p}(0)$ does not
have a special status but when $p \neq p_0 \Rightarrow B \neq 0$, and $B$ is an extra contribution to $\tilde{p}(0)$, see Eq.~\eqref{ansatz}. We now require normalization, that is,
	\begin{align}
	1 & = \sum\limits_{k=0}^N\tilde{p}(k) = A 2^N + B.
	\label{Eq:AB2}
	\end{align}
\end{subequations}
Solving Eqs.~\eqref{Eq:AB} for $A$ and $B$ we find
\begin{subequations}
	\begin{align}
	A &= \frac{1}{2^N + p/p_0 - 1}, \\
	B &= \frac{p/p_0-1}{2^N +p/p_0-1},
	\end{align}
\end{subequations}
yielding
\begin{equation}
\tilde{p}(0) = \frac{p/p_0}{2^N +p/p_0 -1}.
\end{equation}
Having obtained the analytical solutions, the fraction of time the system spends in AF 
for the cMF model is given by 
\begin{equation}\label{TAFC}
1-\tilde{p}(0) = \frac{2^N-1}{2^N +p/p_0 -1}. 
\end{equation}
The time in AF for the cMF model is shown in Fig.~\ref{fig:cmf}.
It is interesting to contrast this result with a simple birth-death process where $p = p_0$. The time in AF is shown for the simple birth-death process and for the cMF calibrated to the CMP model, $p/p_0 = T/\tau = 4.4$, in Table~\ref{cMFtab}. The results indicate that for $N=0$, neither model enters AF. As $N$ is increased, the time in AF is initially much higher in the birth-death process than the cMF, but this difference vanishes as $N$ becomes large. Only when $N \rightarrow \infty$ does the model spend 100\% of the time in AF. Since $N$ must be finite in the CMP model, this indicates that the cMF cannot explain persistent AF.
\vspace{0.5cm}
\begin{table}
\centering
\begin{tabular}{ |c|c|c|c|c|c|c|c|c| } \hline
 $N$     & 0 & 1 & 2 & 3 & 4 & 5 & 10 & $\infty$ \\ \hline 
 $p/p_0=4.4$ & 0 & 0.185 & 0.405 & 0.614 & 0.773 &  0.876 & 0.996 &1 \\ \hline
 $p/p_0 = 1$ & 0 & 0.5 & 0.75 & 0.875 & 0.938 &  0.969 & 0.999 &1 \\ \hline 
\end{tabular}
\caption{Time in AF in the cMF model for different combinations of $p/p_{0}$ and $N$. We observe that for small $N$ the time in AF is significantly higher when the ratio $p/p_{0}$ is small. These differences vanish as $N$ increases.\label{cMFtab}}
\end{table}

\section{Enhanced Mean-Field Model of AF}\label{extendedMF}
The phase diagrams discussed in Fig.~\ref{mfphasediagram} reveal significant differences between the CMP and MF models as the latter underestimates the time in AF. One may assert that this discrepancy stems from a poor replication of the interactions between re-entrant circuits, and in particular from the exclusion of the non-spatial features of critical structures (e.g. particles are assumed to have the same length) from the MF model. In this section, we provide further evidence against this hypothesis by showing that modelling various non-spatial features of critical structures does not mitigate the differences between the CMP and MF models. To do so, we introduce an enhanced version of the MF model (eMF) in which each particle retains the length of the associated critical structure and changes its state at specific time steps, depending on the overall configuration of the system. The purpose of the eMF is to indicate that the non-spatial simplifications in the MF model are not responsible for the discrepancy in the time in AF between the MF model and the CMP model. 

In the eMF model, the system is represented by the state vector $P(t) = (p_{1}(t),\dots,p_{N}(t))$, where $p_{j}(t) \in \{0,1\}, \; j = 1,\dots, N$ is the state of the $j$th particle at time $t$ and $N$ is the number of particles corresponding to the simple critical structures found across the CMP lattice. When $p_{j}(t) = 1$ ($p_{j}(t) = 0$), the $j$th particle is active (inactive) at time $t$. The number of active particles at time $t$ is 
\begin{equation}
N_{a}(t) = \sum\limits_{j = 1}^{N} p_{j}(t).
\end{equation}
In line with the original MF model, the system is in sinus rhythm when $N_{a}(t) = 0$ and in AF when $N_{a}(t) > 0$. The $j$th particle $p_{j}(t)$ can change its state at a specific time $t^{*}_{j}$. We set the first switching time for each particle as $t^{*}_{j} \rightarrow U_{j}$, where $U_{j}, j = 1,\dots,N$ is a uniformly distributed integer random variable in $[1,L]$. This mimics the first planar wave front released from the pacemaker reaching critical regions at different time steps due to their different locations. As soon as the simulation time $t$ matches $t^{*}_{j}$, the $j$th particle changes its state with probability $\epsilon$. Independently of whether the $j$th particle has changed its state or not, its next switching time $t^{*}_{j}$ is updated 
\begin{align} \label{rates2}
t^{*}_{j} & \rightarrow \begin{cases} t^{*}_{j} + \min\limits_{p: p(t + 1) = 1}\ell_p, & \mbox{if } N_{a}(t + 1) > 0 \mbox{ and } p_{j}(t + 1) = 0, \\
t^{*}_{j} + \ell_{j}, & \mbox{if } N_{a}(t + 1) > 0 \mbox{ and } p_{j}(t + 1) = 1,\\
t^{*}_{j} + T, & \mbox{if } N_{a}(t + 1) = 0,
\end{cases}
\end{align}
where $t + 1$ indicates that the update is based on the characteristics of the system observed immediately after the eventual state change of the $j$th particle. When $N_{a}(t + 1) = 0$, the $j$th particle will attempt to switch its state in $T$ time steps. This mimics sinus rhythm in the CMP model where the planar wave front released from the sinus nodes reaches a critical structure every $T$ time steps. When $N_{a}(t + 1) > 0$, particles try to switch their states more frequently. This reflects the intense activity (e.g. number of active nodes per time step) observed in AF episodes occurring in the CMP lattice.

In the eMF model, the length of the shortest active particles dictates the period between two consecutive attempts to activate a dormant region. For instance, the $j$th particle that turns (or remains) off at time $t^{*}_{j} = t$ will attempt to activate again at time 
\begin{equation}
t^{*}_{j} = t^{*}_{j} + \min\limits_{p: p(t + 1) = 1}\ell_p,
\end{equation}
where the final term is the length of the shortest active particles at time $t + 1$. This mimics the fact that in the CMP model the length of a re-entrant circuit determines the frequency at which nodes forming the hosting critical structure emit waves. 

%On the other hand, waves emitted from an active critical structure might terminate other re-entrant circuits. 

%This occurs when meandering waves enter the critical region and excite resting nodes that were about to be reached by the ongoing circuital wavefront. The measurement of the exposure of a re-entrant circuit to external waves poses an arduous challenge as it considers several factors, such as the length of the critical structure, the orientation of the interfering external waves and the state of the nodes they are trying to excite. For the sake of simplicity, the eMF model neglects these aspects by assuming that active particles are only subject to self-termination. In particular, the $j$th particle that becomes (or remains) active at time $t^{*}_{j} = t$ will attempt to deactivate again at time $t^{*}_{j} = t^{*}_{j} + \ell_{p_j}$, where $\ell_{p_j}$ is the length of the $j$th particle.

\begin{figure}
\centering
\includegraphics[width=7.0cm]{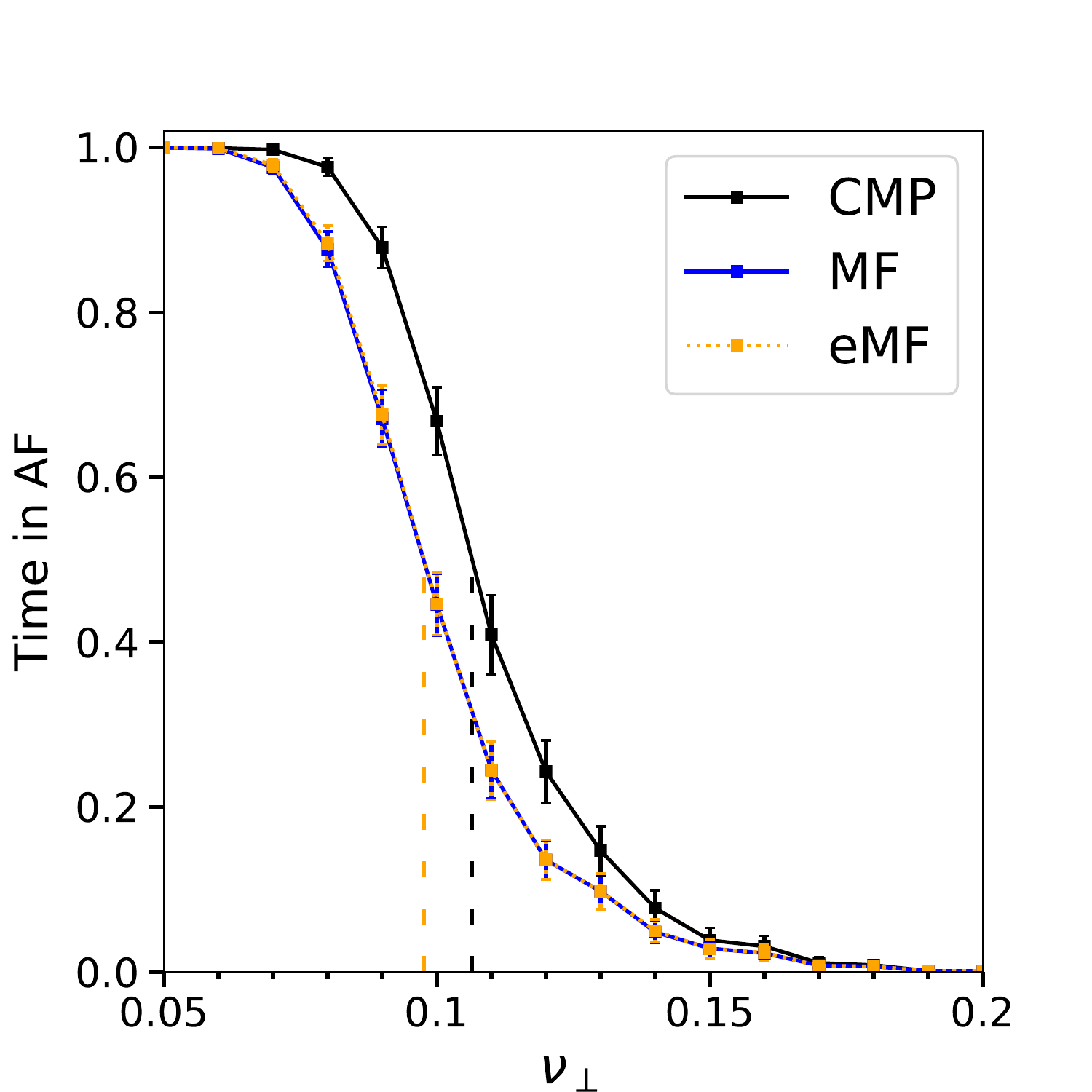}
\caption{Phase diagram of the time in AF as a function of the fraction of transversal connections $\nu_{\perp}$ for the CMP (black), MF (blue) and eMF (orange) models. We use the parameters of the CMP model (i.e.,  $N$, $\epsilon$, and $\langle \ell \rangle$) and the lengths of the tracked simple critical structures (i.e.,  $\ell_{p_1},\dots,\ell_{p_N}$) to calibrate the MF and eMF models. The phase diagrams of the MF and eMF models are perfectly compatible. Both models significantly underestimates the time in AF with respect to the CMP model. These results suggest that the spatial structure of the CMP model is responsible for the excess time the CMP model spends in AF compared to the MF model, and not the non-spatial simplifications of the MF model. Sharp transitions in the time in AF occur around the threshold values $\nu^{*}_{\perp} \approx 0.11$ (CMP, black dashed line), $\nu^{*}_{\perp} \approx 0.09$ (MF, blue dashed line; eMF, orange dashed line).}
\label{emfphasediagram}
\end{figure}

The eMF model enhances the replication of the interactions between simple critical structures by capturing potentially important spatial features that have been excluded from the original simplified MF model. The goal of this framework is to assess the contribution of the non-spatial features of critical structures to the significant discrepancies between the CMP and the MF models, see Fig.~\ref{mfphasediagram}. We find that the phase diagram of the time in AF in the eMF model is perfectly compatible with the one derived from the MF model, see Fig.~\ref{emfphasediagram}. This suggests that adding further layers of complexity to capture every feature of the interactions between simple critical structures is unlikely to reconcile the statistics obtained from the CMP and the MF models. 
%Instead, these results provide additional support to the hypothesis that the higher statistics observed in the CMP phase diagram stems from the additional contribution of different activation mechanisms (e.g.  complex critical structures and re-entrant circuits originating from colliding waves of excitation) that are not considered in the MF model. 

\section{\label{sec:AF_def}CMP model definition of AF \label{sec:ECG}}

\begin{figure}
\centering
\includegraphics[width=\linewidth]{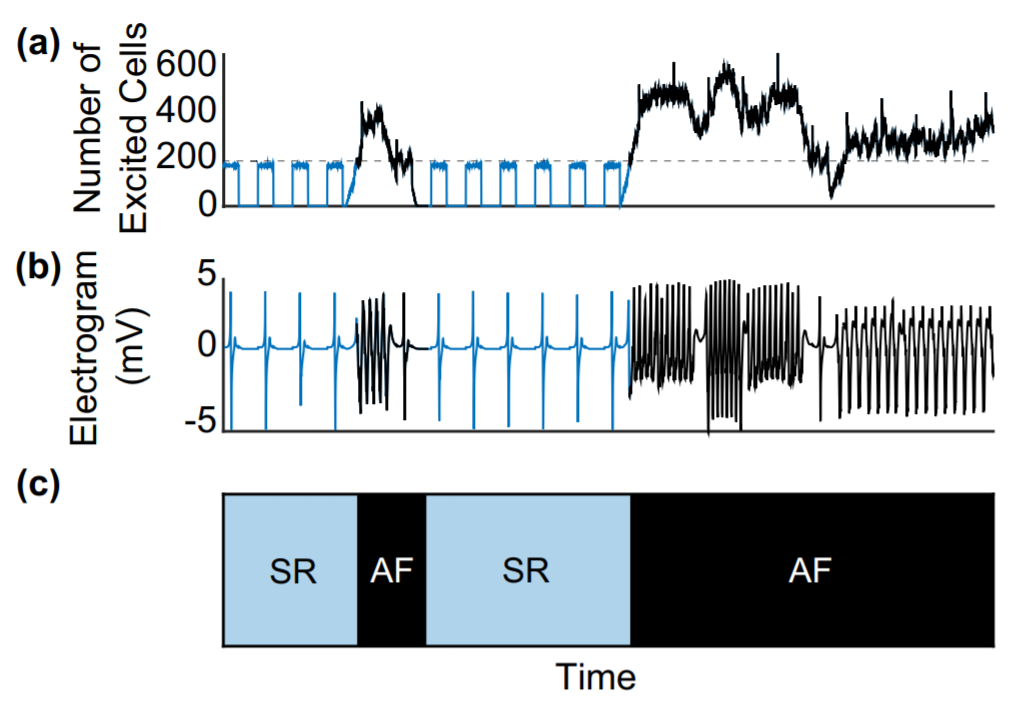}
\caption{The classification of AF in the CMP model. (a) The number of active nodes over time. The dashed line indicates the AF threshold of 220 active nodes. (b) Simulated electrograms derived from the CMP simulations. (c) The classification of whether the model is in sinus rhythm (SR) or AF over time. Blue lines indicate the model is in sinus rhythm, black lines indicate the model is in AF. This figure has been used with permission from \cite{manani2016}.}
\label{fig:clinical_AF}
\end{figure}

The CMP model with $T=220$, $L=200$, $\tau = 50$ is defined to be in AF when the number of active nodes in the model exceeds $1.1L$ (220 nodes). This is a working definition of AF in the CMP model and is not a unique choice. We have previously tested that this definition of AF correlates well with what would be expected from a clinical definition of AF \cite{manani2016}. This is shown in Fig.~\ref{fig:clinical_AF} where we plot (a) the number of active nodes in the model over time, (b) the corresponding simulated electrograms, and (c) the classification of whether the model is in AF or not according to our working definition of AF in the CMP model. The figure shows that the number of active nodes during sinus rhythm follows a regular pattern with only small scale noise around the average number of active nodes. This average falls below the practical definition of AF where we require more than $220$ active nodes. The corresponding electrograms are regular and consistent with sinus rhythm pacing. When a re-entrant circuit forms, the number of active nodes rapidly exceeds the threshold, and rapid, irregular activity is observed in the electrograms. The activation frequency observed is significantly higher than expected in sinus rhythm. This state is classified as being in AF according to our working definition. For more details see \cite{manani2016}. 

Note, we do not explicitly distinguish between atrial tachycardia (AT) and AF. The dynamics in the CMP model are solely based on the formation of re-entrant circuits. These circuits are generally transient and are short lived. In practice, we observe regular rapid pacing in the CMP model when only a single re-entrant circuit has formed. Conversely, rapid irregular pacing is observed when more than one re-entrant circuit forms.

\section{Activation patterns in the CMP Model \label{sec:act_patterns}}

The CMP model is a highly simplified, physics style model of AF, focusing on the initiation and maintenance of micro-anatomical re-entrant circuits. The CMP model does not consider the maintenance of AF from rotors (spiral waves). As a result, the macroscopic activation patterns in the CMP model do not directly reflect what might be observed clinically or in other reaction-diffusion models of fibrillation. 

\begin{figure*}
\centering
\includegraphics[width = 0.8\linewidth]{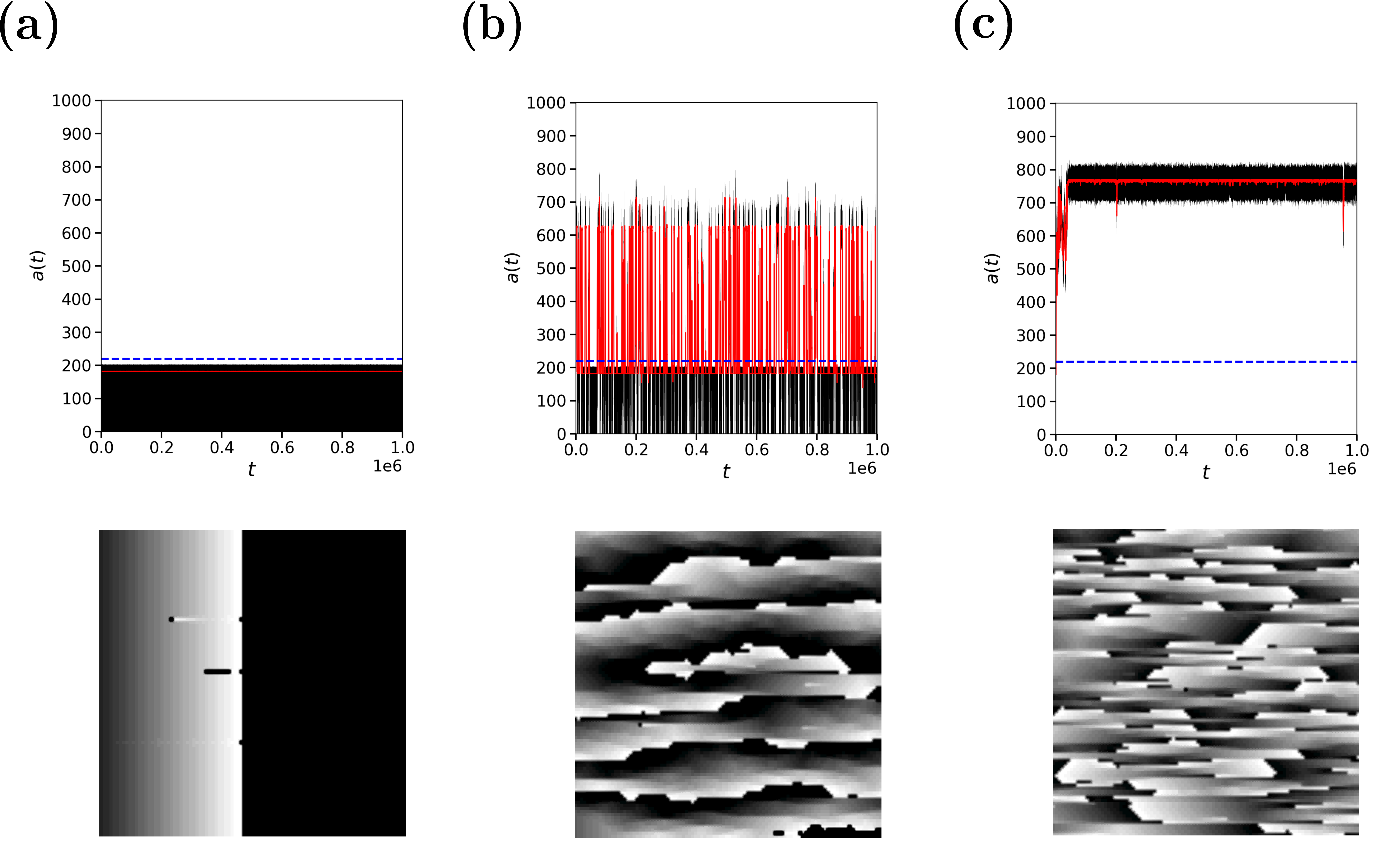}
\caption{The number of active cells per time step, $a(t)$; top row, and the corresponding activation patterns observed in a $100 \times 100$ snapshot of the CMP model, bottom row, in (a) sinus rhythm, (b) paroxysmal AF, and (c) persistent AF. Active nodes are shown in white, refractory in greyscale, and resting in black.}
\label{fig:phenotype}
\end{figure*}

\begin{figure}[h!]
\centering
\includegraphics[width=\linewidth]{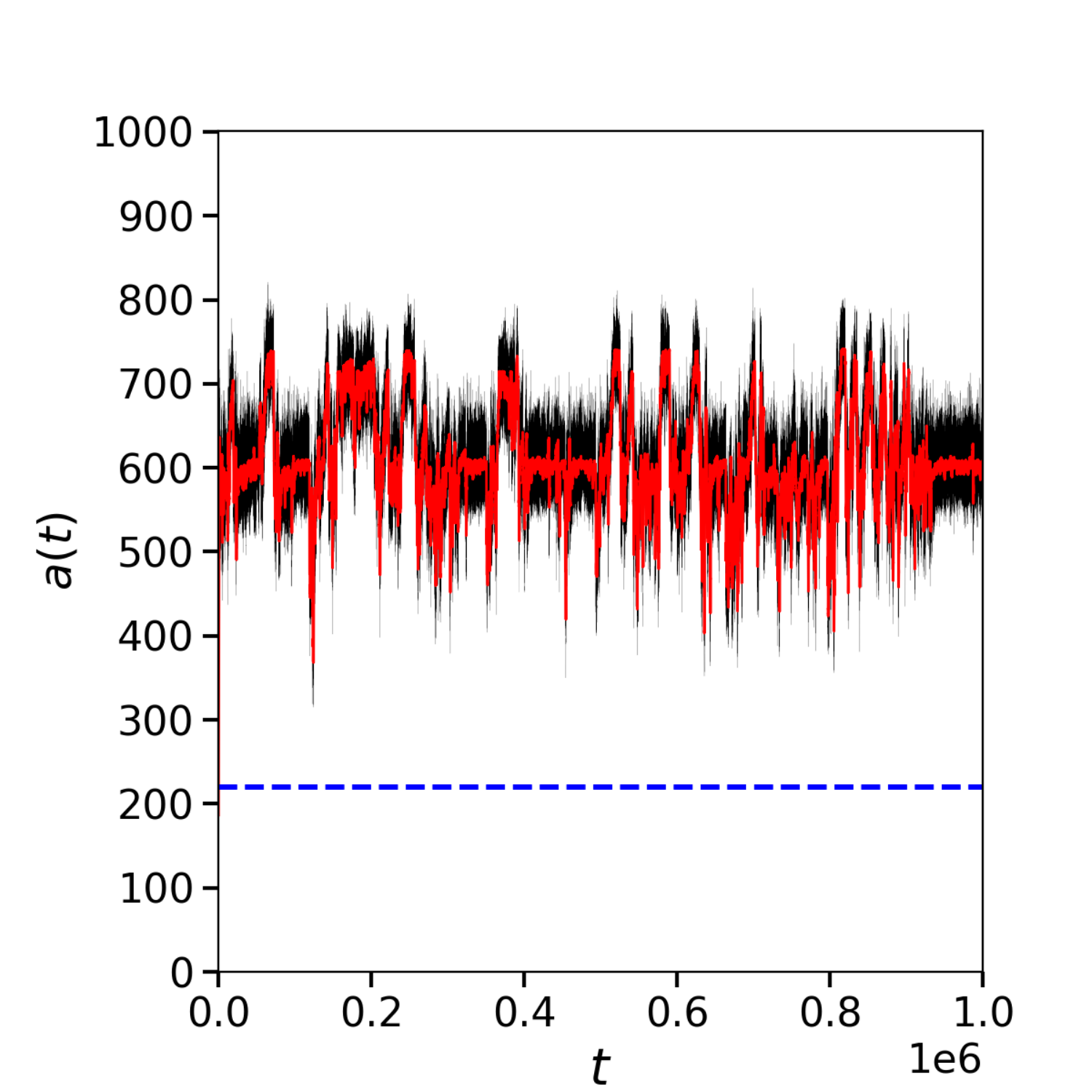}
\caption{An example of persistent AF in the CMP model at $\nu_{\perp} = 0.05$ where the number of active nodes per time step shows significant fluctuations over time. The black line indicates the raw data, with the red line indicating the moving average over a time window of $T=220$. The blue dashed line indicates the threshold above which the CMP model is said to be in persistent AF. }
\label{fig:persistent_unstable}
\end{figure}

Figure~\ref{fig:phenotype} shows typical activation patterns observed in the CMP model and the corresponding trace of the number of active cells. As expected, in sinus rhythm, the number of active cells is constant and falls below the threshold for AF. When the number of active cells exceeds the AF threshold, we can observe a number of different AF phenotypes in the CMP model from paroxysmal to persitent AF. Qualitatively, the activation patterns in paroxysmal and persistent AF do not show major differences, although persistent AF is typically associated with a higher dominant frequency of activation. Interestingly, their is some clinical evidence to suggest that increased dominant frequency predicts an increase in the persistence of AF \cite{Martins2014}.

In Fig.~\ref{modeldynamics}(h), an episode of persistent AF is shown where is the number of active cells is stable over time. The simulation was generated at $\nu_{\perp} = 0.11$, where the average number of simple structures in the CMP model is $N<2$, see appendix~\ref{sec:MF_vs_cCMP}. This suggests that the example shown in Fig.~\ref{modeldynamics}(h) could plausibly be the result of one single stable re-entrant circuit (although not a simple one). As a result, this example may be better thought of as a persistent episode of AT rather than AF. However, many examples of persistent AF are also observed in the CMP model where the number of active cells shows frequent fluctuations, but where the activation remains above the AF threshold, see Fig.~\ref{fig:persistent_unstable}.

\section{Correspondence between the number of simple re-entrant circuits and overall coupling \label{sec:MF_vs_cCMP}}

\begin{figure}
\centering
\includegraphics[width=\linewidth]{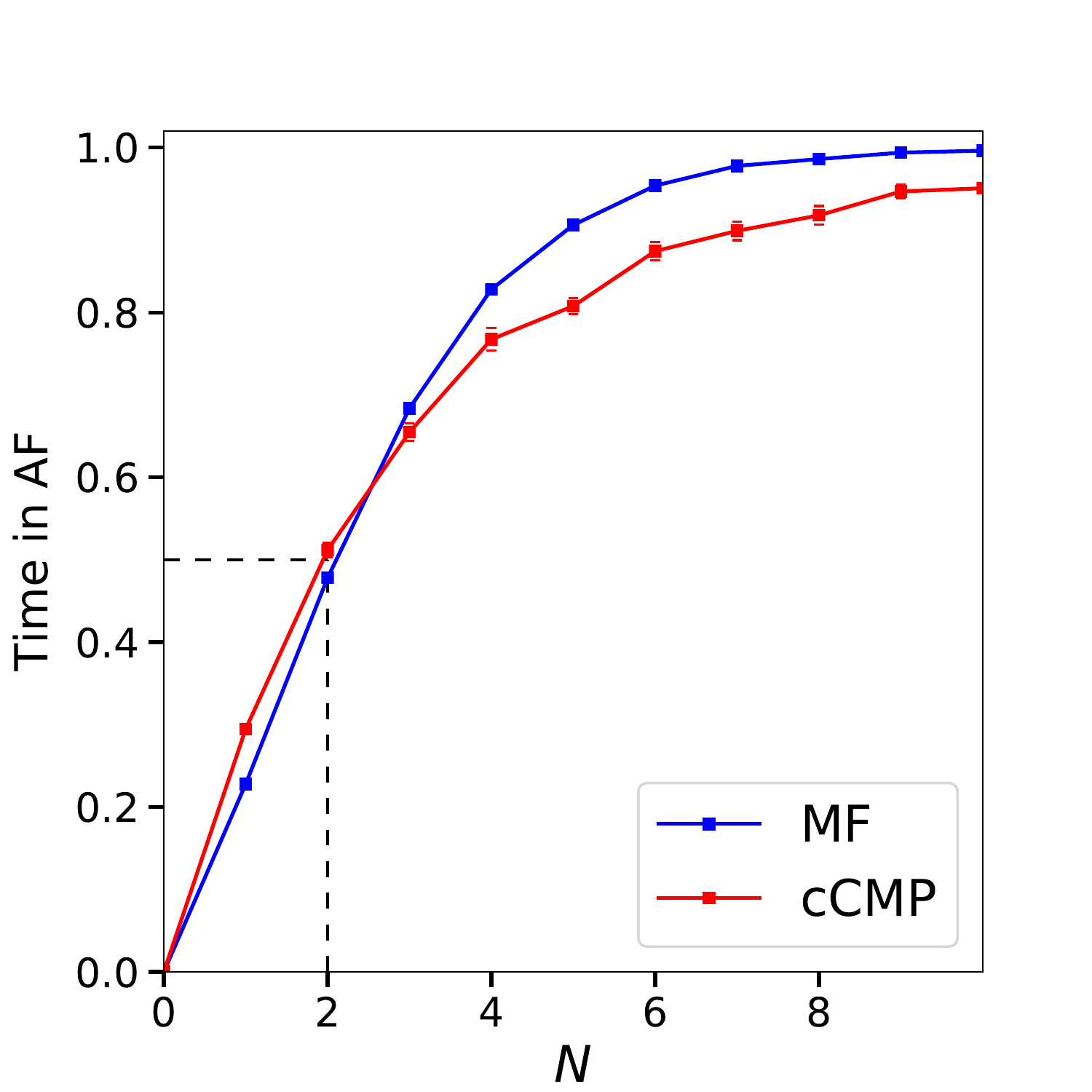}
\caption{The average time in AF for the cCMP (red) and MF (blue) models as a function of (1) the number of tracked simple critical structures $N$ in the cCMP model, or (2) the MF model with the corresponding number of particles.}
\label{fig:time_vs_N}
\end{figure}

\begin{figure}
\centering
\includegraphics[width=\linewidth]{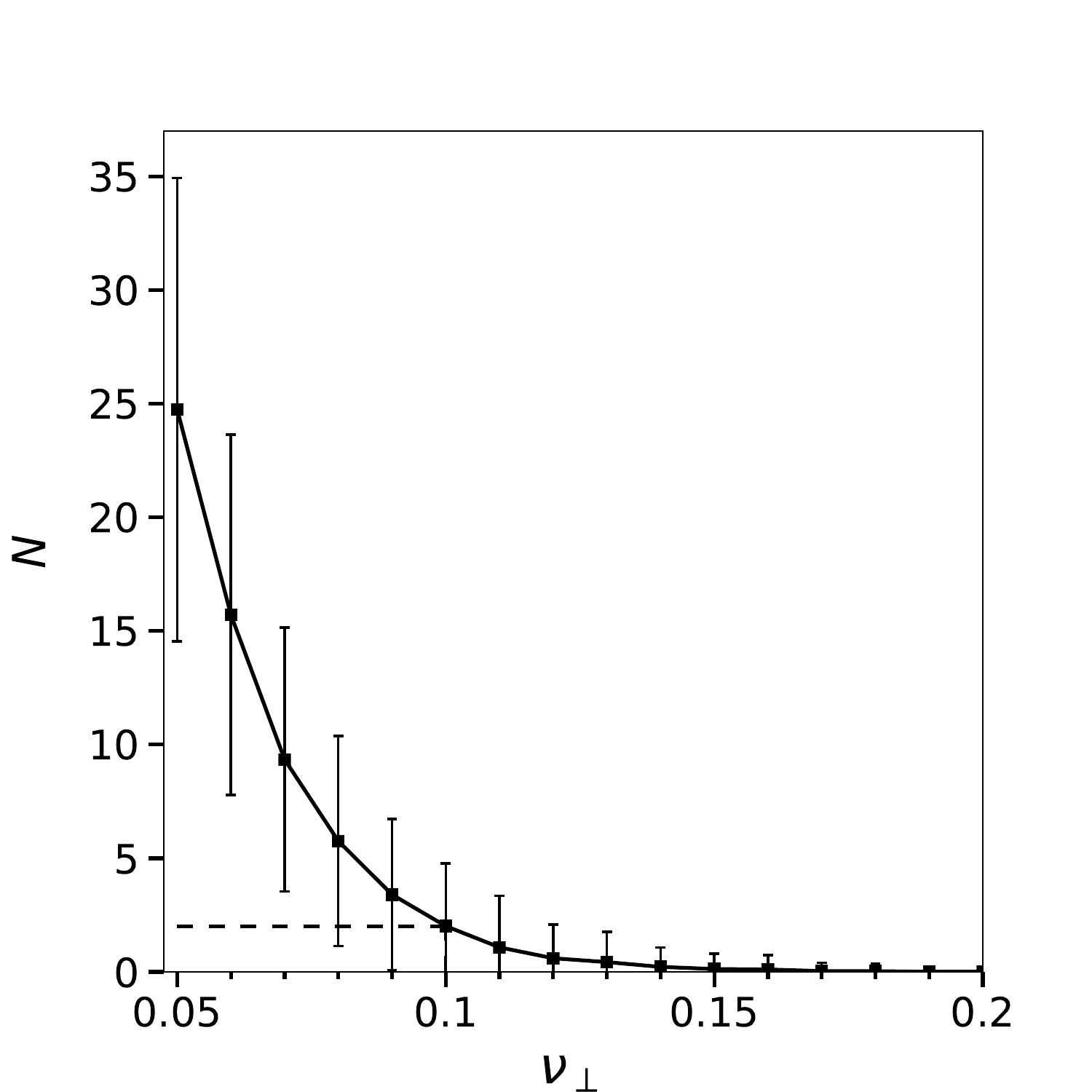}
\caption{The average number of tracked simple critical structures $N$ as a function of the overall coupling value $\nu_{\perp}$ in the cCMP model. The errors bars indicate the 95\% confidence interval calculated over 50 simulations.}
\label{fig:nu_vs_N}
\end{figure}

As discussed in section~\ref{FUNCTIONAL}, it is possible to identify the number of simple critical structures, $N$, regions capable of forming a simple re-entrant circuit, in a given instance of the CMP model. Previously, we have presented the time in AF in the CMP model as a function of the overall coupling $\nu_{\perp}$. 

In Fig.~\ref{fig:time_vs_N} we show the time in AF as a function of the number of critical structures, $N$, identified in the cCMP model where we control the placement of conduction blocking nodes. For each instance of the cCMP model, the corresponding value of $N$ is used to generate a simulation of the MF model. Figure~\ref{fig:time_vs_N} demonstrates that at low $N$ ($N < 2$), the time in AF in the cCMP model slightly exceeds the corresponding value of the MF model. In constrast, the converse is observed at large $N$ ($N > 2$) where the time in AF in the MF model exceeds the corresponding value in the cCMP model. This is a consequence of the spatial elements of the cCMP model which are absent in the MF model. At low $N$, AF episodes last a little longer in the cCMP model than in the MF model (due to slight differences in the activation and deactivation rates of the models). Conversely, at high $N$, the activation (or deactivation) of a particle in the MF model is independent of any other particle in the model, whereas in the cCMP model, an active re-entrant circuit can suppress the activation of other critical structures which are longer than the currently active re-entrant circuit. As a result, the time in AF in the cCMP model is above (below) the MF value for low (high) $N$.

Figure~\ref{fig:nu_vs_N} shows the average number of simple critical structures detected in an instance of the cCMP model as a function of the overall coupling, $\nu_{\perp}$. For Fig.~\ref{fig:nu_vs_N}, the error bars have been chosen to show the 95\% confidence interval of possible $N$ values at a given coupling value. Fig.~\ref{fig:time_vs_N} shows that $N\approx 2$ is the crossover value above (below) which the time in AF is larger (smaller) in the MF model than in the cCMP model. Figure~\ref{fig:nu_vs_N} indicates that $N=2$ corresponds to a coupling value of $\nu_{\perp} \approx 0.1$. Hence, the small difference in the time in AF shown in Fig.~\ref{fullphasediagram} can be understood as being a consequence of the slightly different time in AF values at fixed $N$ indicated in Fig.~\ref{fig:time_vs_N}.

Note, if $N=1$, a single dominant re-entrant circuit drives fibrillation in the CMP model. Hence, at a simplified level, this can be thought of as a form of atrial tachycardia (AT), however, this is very rare at low coupling. When multiple drivers are competing ($N>1$), the activity in the CMP model is better associated with AF.

%When the number of critical structures is small, interactions between individual re-entrant circuits are minimised and act as almost independent structures. In this scenario, the only factors influencing the time in AF is the rate at which these critical structures activate and deactivate. In the cCMP model, a critical structure has a probability $\epsilon$ of activating once every $T$ timesteps. Likewise, once active, the re-entrant circuit has probability $\epsilon$ to terminate once every $\ell$ timesteps, where $\ell$ is the length of the re-entrant circuit. For practical purposes we assume that at small $N$, $\ell \approx \tau$. Conversely, a particle can activate (deactivate) with probability $\epsilon/T$

\section{\label{VIDEO}Videos}
Videos of the different structure types discussed throughout this paper are provided in the supplementary material with appropriate captions \cite{SM}. 

\bibliographystyle{apsrev4-1}
\bibliography{Bibliography}
\end{document}